\documentstyle[12pt,bezier,epsf]{article}
\def\beq{\begin{equation}}
\def\eeq{\end{equation}}

\def\ap#1#2#3 {Ann. Phys. (NY) {\bf#1} (19#2) #3}
\def\apj#1#2#3 {Astrophys. J. {\bf#1} (19#2) #3}
\def\apjl#1#2#3 {Astrophys. J. Lett. {\bf#1} (19#2) #3}
\def\app#1#2#3 {Acta. Phys. Pol. {\bf#1} (19#2) #3}
\def\ar#1#2#3 {Ann. Rev. Nucl. Part. Sci. {\bf#1} (19#2) #3}
\def\cpc#1#2#3 {Computer Phys. Comm. {\bf#1} (19#2) #3}
\def\err#1#2#3 {{\it Erratum} {\bf#1} (19#2) #3}
\def\ib#1#2#3 {{\it ibid.} {\bf#1} (19#2) #3}
\def\jmp#1#2#3 {J. Math. Phys. {\bf#1} (19#2) #3}
\def\ijmp#1#2#3 {Int. J. Mod. Phys. {\bf#1} (19#2) #3}
\def\jetp#1#2#3 {JETP Lett. {\bf#1} (19#2) #3}
\def\jpg#1#2#3 {J. Phys. G. {\bf#1} (19#2) #3}
\def\mpl#1#2#3 {Mod. Phys. Lett. {\bf#1} (19#2) #3}
\def\nat#1#2#3 {Nature (London) {\bf#1} (19#2) #3}
\def\nc#1#2#3 {Nuovo Cim. {\bf#1} (19#2) #3}
\def\nim#1#2#3 {Nucl. Instr. Meth. {\bf#1} (19#2) #3}
\def\np#1#2#3 {Nucl. Phys. {\bf#1} (19#2) #3}
\def\pcps#1#2#3 {Proc. Cam. Phil. Soc. {\bf#1} (#2) #3}
\def\pl#1#2#3 {Phys. Lett. {\bf#1} (19#2) #3}
\def\prep#1#2#3 {Phys. Rep. {\bf#1} (19#2) #3}
\def\prev#1#2#3 {Phys. Rev. {\bf#1} (19#2) #3}
\def\prl#1#2#3 {Phys. Rev. Lett. {\bf#1} (19#2) #3}
\def\prs#1#2#3 {Proc. Roy. Soc. {\bf#1} (19#2) #3}
\def\ptp#1#2#3 {Prog. Th. Phys. {\bf#1} (19#2) #3}
\def\ps#1#2#3 {Physica Scripta {\bf#1} (19#2) #3}
\def\rmp#1#2#3 {Rev. Mod. Phys. {\bf#1} (19#2) #3}
\def\rpp#1#2#3 {Rep. Prog. Phys. {\bf#1} (19#2) #3}
\def\sjnp#1#2#3 {Sov. J. Nucl. Phys. {\bf#1} (19#2) #3}
\def\spj#1#2#3 {Sov. Phys. JEPT {\bf#1} (19#2) #3}
\def\spu#1#2#3 {Sov. Phys.-Usp. {\bf#1} (19#2) #3}
\def\zp#1#2#3 {Zeit. Phys. {\bf#1} (19#2) #3}
\begin{document}
\begin{titlepage}
\begin{center}
{\Large \bf Theoretical Physics Institute \\
University of Minnesota \\}  \end{center}
\vspace{0.15in}
\begin{flushright}
TPI-MINN-94/33-T \\
UMN-TH-1314-94 \\
September 1994
\end{flushright}
\vspace{0.2in}
\begin{center}
{\Large \bf Non-perturbative methods\\ }
\vspace{0.15in}
{\bf M.B. Voloshin  \\ }
Theoretical Physics Institute, University of Minnesota,
Minneapolis, MN 55455 \\
and \\
Institute of Theoretical and Experimental Physics,
Moscow, 117259 \\
\vspace{0.4in}
\end{center}

\abstract{The current theoretical understanding of processes involving many
weakly interacting bosons in the Standard Model and in model theories is
discussed. In particular, such processes are associated with the baryon and
lepton number violation in the Standard Model. The most interesting domain
where the multiplicity of bosons is larger than the inverse of small
coupling constant is beyond the scope of perturbation theory and requires a
non-perturbative analysis.}

\vspace{0.4in}
\begin{center}
{\it Plenary talk presented at the 27th International Conference on High
Energy Physics \\
Glasgow, Scotland, 20--27 July 1994}
\end{center}

\end{titlepage}

\section{Introduction}

The vastness of the field of non-perturbative methods in high-energy physics
inevitably compels me to focus on a specific topic among those which
attract a considerable interest and where a non-trivial development is
likely in the near future. One such topic, which also has dominated the
parallel session on non-perturbative methods, is related to multiboson
phenomena in the electroweak physics and, more generally, in models with
weak coupling. These phenomena are interesting because of two basic reasons.
One is that it is with multiboson processes is associated the violation of
the sum of the baryon (B) and the lepton (L) numbers in the Standard
Model\cite{hooft}.  Therefore such processes determine the evolution of
(B+L) at high temperature in the early universe\cite{krs}. Also as initially
envisioned in early works\cite{christ,ag} and indicated by specific
calculations\cite{ringwald,espinosa} the processes with (B+L) violation and
production of many electroweak bosons might be in principle observable in
high energy collisions in the multi-TeV energy range. The other,
theoretical, reason is related to the old-standing problem of the factorial
divergence of perturbation theory series, which dates back to the work of
Dyson\cite{dyson}. This problem looks to be a matter of a purely theoretical
concern as long as the quantities under discussion are such that they appear
at low orders, like the anomalous magnetic moment. For such quantities the
inability in principle to find the exact result by the perturbative
expansion, though disappointing, does not prevent from calculating in few
first orders with an accuracy required by practical measurements or greater.
However the problem of the inherent divergence of perturbative series
becomes quite acute as soon as one considers processes at energies such that
a large number of interacting particles can be produced, i.e. the processes
which occur starting only from a high order of the perturbation theory,
where the expansion becomes unreliable.

At present there is a general understanding that multiparticle electroweak
processes with many bosons both in the initial and the final state ({\em
many} $\to$ {\em many} scattering), including those with (B+L) violation,
are not suppressed at high temperature and thus they indeed determine the
(B+L) history of the universe. On the other hand the understanding of the
processes, in which many bosons are produced by two or few initial particles
with high energy ({\em few} $\to$ {\em many} scattering), is far from
complete and there are arguments both pro and con the idea that at
sufficiently high multiplicity of final particles such processes can have an
observable cross section.

The contribution of the {\em many} $\to$ {\em many} scattering at high
temperature is described within the WKB technique by expansion around
non-trivial classical solutions of the field equations:
sphalerons\cite{manton,km} and, more generally, periodic
instantons\cite{krt}. It is also believed that the {\em few} $\to$ {\em
many} scattering can be fully described by applying a WKB technique using
special classical configurations of the field. However, this is still a
conjecture, and a specific method of a full WKB analysis of the latter
scattering has yet to be developed.

\section{(B+L) violating electroweak processes}

As is known\cite{hooft}, the electroweak interaction in the Standard Model
does not conserve the sum of the baryon and the lepton numbers as a result
of the triangle anomaly. The amount of the (B+L) violation in a
process is determined by the change of the winding number $N_{CS}$ of the
electroweak gauge fields:
\beq
\Delta \, (B+L)=6\,\Delta\, N_{CS}~~.
\label{anom}
\eeq
However, changing the winding number by one or several units requires
presence (at least in the intermediate state) of the $W$ and $Z$ field
configurations with energy of order $m_W/\alpha_W$. This is usually
illustrated by the sketch of the dependence of minimal energy of the field
with a given $N_{CS}$ as a function of $N_{CS}$ shown in figure 1, and the
field configuration, corresponding to the top of the barrier, is the
so-called sphaleron\cite{manton} with energy $E_{\rm Sp}$ about 10 TeV
\cite{km}.

\thicklines
\unitlength=1.00mm
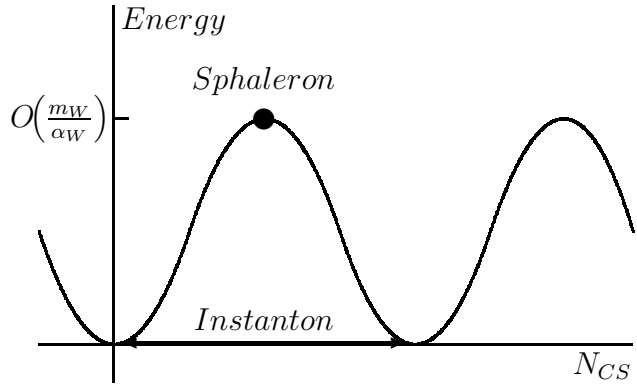
\begin{figure} \begin{center}
\begin{picture}(84.00,55.00)(0,48.00)
\bezier{252}(25.00,70.00)(35.00,100.00)(45.00,70.00)
\bezier{252}(25.00,70.00)(15.00,40.00)(5.00,70.00)
\bezier{252}(45.00,70.00)(55.00,40.00)(65.00,70.00)
\bezier{252}(65.00,70.00)(75.00,100.00)(84.00,70.00)
\put(5.00,55.00){\line(1,0){79.00}}
\put(15.00,50.00){\line(0,1){50.00}}
\put(15.00,85.00){\line(1,0){2.00}}
\put(35.00,55.20){\vector(1,0){19.00}}
\put(35.00,55.20){\vector(-1,0){19.00}}
\put(35.00,85.00){\circle*{2.83}}
\put(16.00,98.00){\makebox(0,0)[lc]{$Energy$}}
\put(80.00,54.00){\makebox(0,0)[ct]{${ N_{CS}}$}}
\put(14.50,85.00){\makebox(0,0)[rc]{$O\! \! \left ( {{m_W} \over {\alpha_W}}
\right )$}}
\put(35.00,89.00){\makebox(0,0)[cb]{$Sphaleron$}}
\put(35.00,57.00){\makebox(0,0)[cb]{$Instanton$}}
\end{picture}
\caption{The dependence of minimal energy of the electroweak gauge field on
its winding number.}
\end{center} \end{figure}

If the energy $E$ available in a process is much less than $E_{\rm Sp}$ then
the only way in which (B+L) can be violated is due to quantum tunneling,
which at $E =0$ is described by the instanton\cite{bpst} solution to the
Euclidean field equations, whose action is $S_{\rm i}=2\pi/\alpha_W$. The
amplitude of such process then contains the WKB tunneling factor $\exp
(-2\pi/\alpha_W) \sim 10^{-80}$, which thus makes the process unobservable
by any practical measure.

The plot in the figure 1 however invites the suggestion that once the
available energy is close to or larger than $E_{\rm Sp}$ the suppression of
(B+L) violating processes should weaken or disappear altogether. The two
relevant situations where large energy is available in individual processes
are high temperatures and high-energy particle collisions.

\subsection{(B+L) violation at high temperature}

As first realized by Kuzmin, Rubakov and Shaposhnikov\cite{krs} the
rate with which the system traverses the sphaleron barrier in thermal
equilibrium at a temperature $T < E_{\rm Sp}$ is determined by the
Boltzmann factor $\exp(-E_{\rm Sp}(T)/T)$ and may become unsuppressed at
temperatures larger than $E_{\rm Sp}$. The dependence of $E_{\rm Sp}$ on the
temperature arises through the temperature dependence of the vacuum
expectation value of the Higgs field, which sets the electroweak energy
scale. In particular the v.e.v. vanishes at the phase transition temperature
$T_c$, above which the electroweak symmetry is restored. Thus at $T > T_c$
the sphaleron barrier is absent and the (B+L) violating processes may go
without an exponential suppression. As a result [10, 12 - 14] the
rate of change of the (B+L) density is given by
\beq
\Gamma_{\Delta(B+L)}=
\left\{ \begin{array}{ll} C_1\, \exp \left ( -E_{\rm Sp}(T)/T \right) &
\mbox{if $T < T_c$} \\ C_2\, \alpha_W^4 T                          &
\mbox{if $T > T_c$} \end{array} \right.
\label{trate}
\eeq
where $C_1$ and $C_2$ are constants.

In particular, the prefactor $C_1$ for the `low' temperature rate is
determined by fluctuations of the fields near the sphaleron configuration.
The initial calculations [15 - 17], which considered only the
contribution of bosonic fluctuations, were most recently reanalyzed and
extended\cite{901} to include also the fermionic determinant. The
contribution of the heavy top quark is found to significantly suppress the
prefactor $C_1$, which enables to somewhat relax the upper bound on the mass
of Higgs boson in the minimal Standard Model, following from the
requirement\cite{krt,bs,shaposhnikov} that the (B+L) violating processes in
the early universe immediately after the electroweak phase transition (i.e.
just below $T_c$) do not wash out completely the baryon asymmetry,
independently of the mechanism by which it was created before or during the
phase transition. Using their result for $C_1$ and $m_t=174 \, {\rm GeV}$,
Diakonov {\it et.al.} \cite{901} find this upper bound to be $m_H < 66\,
{\rm GeV}$, which is only slightly higher than the lower bound $m_H > 58.4
{\rm GeV}$ \cite{pdg} from a direct search at LEP. Therefore an improvement
in the experimental search can either find the Higgs boson or close the gap
of compatibility of the minimal model with the observed baryon asymmetry of
the universe.

The shape of the sphaleron barrier in the presence of heavy top and the
evolution of the energy levels of a heavy fermion were considered in
detail in the contributed papers \cite{925} and \cite{817} respectively.
Multi-sphaleron configurations are considered in \cite{771} and electroweak
strings, viewed as ``stretched sphalerons" in \cite{938}. However the role
of the latter configurations in thermal equilibrium is yet to be clarified.

\subsection{(B+L) violation in high-energy collisions}

The sphaleron energy scale $E_{\rm Sp}$ is within (hopefully) reachable
energies at prospective colliders. Therefore a most intriguing question
arises as whether the exponential suppression of the (B+L) violating
processes vanishes at an energy of order $E_{\rm Sp}$ in collision of two
leptons or quarks.The difference between the high-temperature (B+L)
violation and the processes induced by just two or few energetic particles
is that in the former case the dominant contribution to the rate
comes\cite{am2} from processes in the thermal bath, in which many soft
particles with total energy $E\; ^>_\sim \; E_{\rm Sp}$ scatter into a final
state of also soft particles with different (B+L), while in the latter case
a coupling between the hard initial particles and soft modes of the field
with a non-trivial topology is required.

Following the conjecture\cite{am2} that an enhancement of the cross section
of (B+L) violating scattering may be associated with multiparticle final
states, Ringwald\cite{ringwald} and Espinosa\cite{espinosa} pursued a
calculation of a generic instanton-induced process of the type
\beq
{\overline f}+{\overline f} \to 10\,f+n_W\, W+n_H\,H~,
\label{genproc}
\eeq
where $n_W$ ($n_H$) is the multiplicity of produced gauge (Higgs) bosons and
$f$ stands for a quark or a lepton. (The presence in the instanton-induced
scattering of twelve fermions:  nine quarks and three leptons, one from each
electroweak doublet is mandated by the anomaly condition in eq.(\ref{anom}),
i.e. by the number of fermionic zero modes of an instanton.) The amplitude
of the scattering (\ref{genproc}) was found to factorially depend on the
multiplicity of bosons: $A \sim n_W!\, n_H!\, \exp(-S_{\rm i})$, which lead
to the argument\cite{mvv} that the factorial enhancement may beat the
exponential suppression at $n_{W,H} > O(1/\alpha_W)$ i.e. at energy
larger than $O\!(E_{Sp})$. In fact the growth with energy of the total
cross section for (B+L) violating processes, observed in the early
calculations\cite{ringwald,espinosa,mvv},
suggested\cite{mvv} that this cross section may become strong: reach its
unitarity limit at energies in the multi-TeV range.

\subsubsection{``Holy grail" function.}
By quite general scaling arguments\cite{amat,krt1} the total cross-section
of instanton-induced scattering should obey the scaling behavior
\beq
\sigma_{\Delta (B+L)}^{tot} \sim \exp \left [ -{{4\pi} \over {\alpha_W}}\,
F \left (  {E \over {E_0}} \right ) \right]~,
\label{holy}
\eeq
where $E_0 \sim E_{\rm Sp} \sim m_W/\alpha_W$. The function $F(\epsilon)$ is
often termed as ``holy grail" function. At $\epsilon=0$ one has $F(0)=1$,
while the initial enhancement\cite{ringwald,espinosa} of the cross section
due to opening multi-boson channels corresponds\cite{viz,porrati} to the
first non-trivial term in the expansion in $\epsilon$: $F(\epsilon)=1-{9
\over 8} \epsilon^{4/3}+\ldots$,  where $\epsilon=E/E_0$ with
$E_0=\sqrt{6}\,\pi\,m_W/\alpha_W \approx 18\,{\rm TeV}$.
The expansion in fact goes in powers of
$\epsilon^{2/3}$, and by now two next terms are
known[30 - 34]:
\beq
F(\epsilon)=1-{9 \over 8} \epsilon^{4/3} +{9 \over {16}} \epsilon^2 +
{3 \over {32}} \left( 4- 3 \, {{m_H^2} \over {m_W^2}} \right)\,
\epsilon^{8/3} \, \ln \epsilon +\ldots
\label{hgf}
\eeq
The latter two terms are determined by interaction between soft final
particles.  Starting from the term of order $\epsilon^{10/3}$ the ``holy
grail" function is also contributed by interactions between hard initial and
soft final particles and by interaction between the initial hard
particles[35 - 37].

Unfortunately, any finite number of terms in the expansion of $F(\epsilon)$
does not allow to assert the behavior of the function at finite $\epsilon
\sim O\!(1)$. Therefore it is not known yet, whether the function
$F(\epsilon)$: \\
{\it i}) goes to zero at finite $\epsilon$ (finite energy),\\
{\it ii}) goes to zero as $\epsilon \to \infty$, or         \\
{\it iii}) is bounded from below by a positive value.           \\
Certainly, the most interesting phenomenologically is the first possibility,
since then the cross section with (B+L) violation and multiboson production
becomes observably large at a finite energy, while the most discouraging
would be the last case, since then the cross section would stay
exponentially suppressed at all energies.

The possibility {\it iii} was advocated[38 - 40] in terms
of the so-called ``premature unitarization"\cite{mash}. The argument is
based on considering the interplay in the $s$-channel unitarity of the
processes {\em few} $\to$ {\em many} and {\em many} $\to$ {\em many}. The
former processes are argued to be still weak (exponentially suppressed) when
the processes {\em many} $\to$ {\em many} are at the unitarity limit, which
effectively shuts off the further growth of the {\em few} $\to$ {\em many}
cross section.

\unitlength=0.9mm
\begin{figure} \begin{center}
\begin{picture}(84.00,44.00)(0,4)
\put(10.00,15.00){\circle{10.00}}
\put(30.00,15.00){\circle{10.00}}
\put(50.00,15.00){\circle{10.00}}
\put(70.00,15.00){\circle{10.00}}
\bezier{48}(15.00,16.00)(20.00,19.00)(25.00,16.00)
\bezier{68}(14.00,18.00)(20.00,24.00)(26.00,18.00)
\bezier{48}(15.00,14.00)(20.00,11.00)(25.00,14.00)
\bezier{68}(14.00,12.00)(20.00,6.00)(26.00,12.00)
\bezier{48}(35.00,16.00)(40.00,19.00)(45.00,16.00)
\bezier{68}(34.00,18.00)(40.00,24.00)(46.00,18.00)
\bezier{48}(35.00,14.00)(40.00,11.00)(45.00,14.00)
\bezier{68}(34.00,12.00)(40.00,6.00)(46.00,12.00)
\bezier{48}(55.00,16.00)(60.00,19.00)(65.00,16.00)
\bezier{68}(54.00,18.00)(60.00,24.00)(66.00,18.00)
\bezier{48}(55.00,14.00)(60.00,11.00)(65.00,14.00)
\bezier{68}(54.00,12.00)(60.00,6.00)(66.00,12.00)
\put(1.00,20.00){\line(5,-2){5.00}}
\put(1.00,10.00){\line(5,2){5.00}}
\put(79.00,20.00){\line(-5,-2){5.00}}
\put(79.00,10.00){\line(-5,2){5.00}}
\put(10.00,15.00){\makebox(0,0)[cc]{$I$}}
\put(30.00,15.00){\makebox(0,0)[cc]{$\overline I$}}
\put(50.00,15.00){\makebox(0,0)[cc]{$I$}}
\put(70.00,15.00){\makebox(0,0)[cc]{$\overline I$}}
\put(10.00,22.00){\makebox(0,0)[cb]{$e^{-S_{\rm i}}$}}
\put(20.00,23.00){\makebox(0,0)[cb]{$B(E)$}}
\put(30.00,22.00){\makebox(0,0)[cb]{$e^{-S_{\rm i}}$}}
\put(40.00,23.00){\makebox(0,0)[cb]{$B(E)$}}
\put(50.00,22.00){\makebox(0,0)[cb]{$e^{-S_{\rm i}}$}}
\put(60.00,23.00){\makebox(0,0)[cb]{$B(E)$}}
\put(70.00,22.00){\makebox(0,0)[cb]{$e^{-S_{\rm i}}$}}
\put(20.00,15.00){\makebox(0,0)[cb]{{\bf \ldots}}}
\put(40.00,15.00){\makebox(0,0)[cb]{{\bf \ldots}}}
\put(60.00,15.00){\makebox(0,0)[cb]{{\bf \ldots}}}
\put(35.00,35.00){\circle{10.00}}
\put(55.00,35.00){\circle{10.00}}
\bezier{48}(40.00,36.00)(45.00,39.00)(50.00,36.00)
\bezier{68}(39.00,38.00)(45.00,44.00)(51.00,38.00)
\bezier{48}(40.00,34.00)(45.00,31.00)(50.00,34.00)
\bezier{68}(39.00,32.00)(45.00,26.00)(51.00,32.00)
\put(26.00,40.00){\line(5,-2){5.00}}
\put(26.00,30.00){\line(5,2){5.00}}
\put(35.00,35.00){\makebox(0,0)[cc]{$I$}}
\put(55.00,35.00){\makebox(0,0)[cc]{$\overline I$}}
\put(35.00,42.00){\makebox(0,0)[cb]{$e^{-S_{\rm i}}$}}
\put(45.00,43.00){\makebox(0,0)[cb]{$B(E)$}}
\put(55.00,42.00){\makebox(0,0)[cb]{$e^{-S_{\rm i}}$}}
\put(45.00,35.00){\makebox(0,0)[cb]{{\bf \ldots}}}
\put(64.00,40.00){\line(-5,-2){5.00}}
\put(64.00,30.00){\line(-5,2){5.00}}
\put(70.00,35.00){\makebox(0,0)[cc]{\large +}}
\put(78.00,15.00){\makebox(0,0)[lc]{{\large + \ldots}}}
\put(23.00,35.00){\makebox(0,0)[rc]{{\Large $\sigma^{tot}=$ Im}}}
\put(26.00,43.00){\line(-1,0){2.00}}
\put(24.00,43.00){\line(0,-1){16.00}}
\put(24.00,27.00){\line(1,0){2.00}}
\put(88.00,23.00){\line(1,0){2.00}}
\put(90.00,23.00){\line(0,-1){16.00}}
\put(90.00,7.00){\line(-1,0){2.00}}
\end{picture}
\caption{Unitarization in the $s$-channel according to the model[40] of
``premature unitarization".}
\end{center} \end{figure}
\unitlength 1.0mm

A somewhat simplified picture of this behavior is shown in
figure 2, where the total cross section is represented as imaginary part of
a $2 \to 2$ forward scattering amplitude through an instanton ($I$) -
antiinstanton (${\overline I}$) configuration. According to the model of
``premature unitarization"\cite{mash} the total amplitude is given by
summation over instanton - antiinstanton chains iterated in the $s$-channel,
i.e. where all the total energy flows through the additional
(anti)instantons. Each additional $I-{\overline I}$ pair brings in the
factor $e^{-2\,S_{\rm i}} \, B(E)$, where the ``bond function" $B(E)$ is the
multi-boson enhancement of the one-instanton-induced cross section observed
in \cite{ringwald,espinosa,mvv}. The summation over the $I-{\overline I}$
chains (figure 2) gives
\beq
\sigma^{tot} \sim e^{-\,S_{\rm i}}\,{\rm Im} \left [ {{B(E) e^{-\,S_{\rm
i}}} \over {1+ \eta \left ( B(E)e^{-\,S_{\rm i}} \right )^2 }} \right ]~,
\label{pu}
\eeq
where $\eta = O(1)$ is a rescattering factor. If given by eq.(\ref{pu}), the
cross section reaches its maximum when $B(E) e^{-\,S_{\rm i}} = O(1)$ and
its value at the maximum is of order $e^{-\,S_{\rm i}}$, which corresponds
to the lower bound of $1/2$ for the function $F(\epsilon)$. The presented
reasoning is however oversimplified: it assumes that all the
(anti)instantons in the chains have same fixed size. Relaxing this
assumption leads\cite{mash} to a lower bound for $F(\epsilon)$, which is
generally different from $1/2$.

At still higher energies the formula (\ref{pu}) gives a falling cross
section. However this regime is unphysical: initial particle can shake off
energy by emitting one or few hard bosons, so that the energy in the
collision gets back to the one corresponding to the maximum. (Emission of
hard bosons suppresses the cross section by a few powers of the coupling
constant, while the gain in the non-perturbative amplitude is exponential.)
If indeed the ``holy grail" function has a minimum at some energy, this
would imply that above that energy the process can not be described by
semiclassical methods, since emission of hard quanta becomes essential.

It turns out however, that the ``premature unitarization" and thus the
simple picture of the $s$-channel iteration of instanton-antiinstanton
correlations is not mandatory and apparently depends on specifics of the
theory. The known examples of simplified models, where the ``holy grail"
function is indeed bounded from below by 1/2 are the Quantum Mechanical
problem with a double well potential\cite{cortik,dp2} and the soft
contribution to the scattering through a bounce\cite{coleman,cc} in a (1+1)
dimensional model of one real field with metastable vacuum\cite{mv2}. (It
has been pointed out\cite{kiselev,rst} that in the latter model there is
also a hard contribution to the bounce-induced scattering, for which the
``holy grail" function goes to zero at the analog of the sphaleron energy.)
Another example, where the ``holy grail" function is bounded by a value,
smaller than 1/2, namely 0.160, is the problem of catalysis of false vacuum
decay in (3+1) dimensions by collision of two (or few) particles\cite{mv3}.
In this problem the semiclassical probability reaches maximum at the top of
the energy barrier.

\subsubsection{Rubakov - Tinyakov approach.}
The main difficulty in developing a semiclassical approach to the {\em few}
$\to$ {\em many} scattering is the presence of hard quanta in the initial
state, which state is thus not a semiclassical one. It has been
suggested\cite{rt,tinyakov} to circumvent this difficulty by considering a
scattering, where a finite small number of particles in the initial state is
replaced by $n_{i(nitial)}=\nu/g^2$, where $g$ is the coupling constant in
the theory and $\nu$ is a parameter. For a finite $\nu$ the initial state of
this kind can be treated semiclassically, and in the end the limit of the
probability at $\nu \to 0$, or $\nu \to {\rm const}/g^2$ is to be considered
in order to relate to the process {\em few} $\to$ {\em many}. Within such
setting the ``holy grail" function depends on $\nu$: $F(\epsilon, \nu)$ and
it is conjectured that its limit at $\nu \to 0$ is smooth, which conjecture
is supported by high-order perturbative calculations around the
instanton\cite{mueller2}. The central point of this approach is that the
function $F(\epsilon, \nu)$ is determined from a solution to a well-defined
boundary value problem\cite{rst2} for classical field equations, although in
essentially complex time.

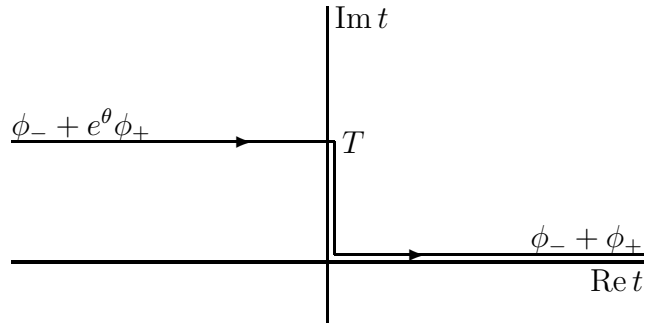
\begin{figure} \begin{center}
\begin{picture}(84.00,49.00)
\put(0.00,15.00){\line(1,0){84.00}}
\put(42.00,7.00){\line(0,1){42.00}}
\put(0.00,31.00){\vector(1,0){32.00}}
\put(32.00,31.00){\line(1,0){11.00}}
\put(43.00,31.00){\line(0,-1){15.00}}
\put(43.00,16.00){\vector(1,0){12.00}}
\put(55.00,16.00){\line(1,0){29.00}}
\put(44.00,31.00){\makebox(0,0)[lc]{$T$}}
\put(43.00,49.00){\makebox(0,0)[lt]{Im$\,t$}}
\put(84.00,14.00){\makebox(0,0)[rt]{Re$\,t$}}
\put(0.00,32.00){\makebox(0,0)[lb]{$\phi_-+e^\theta \phi_+$}}
\put(84.00,17.00){\makebox(0,0)[rb]{$\phi_-+\phi_+$}}
\end{picture}
\caption{Contour in the complex time plane and boundary conditions for a
classical solution, describing the scattering of a semiclassical initial
state into multiparticle final states[52].}
\end{center} \end{figure}

The classical solution that describes the path of largest probability in a
model with one real field $\phi$ is evolving along the contour in the
complex time plane shown in figure 3.  At Re$\,t \to +\infty$ the solution
is required to be real, thus its momentum components should be of the form
$\phi({\bf k})=b_{\bf k}\, e^{-i\,\omega_{k} t}+ b_{-\bf k}^*\,
e^{i\,\omega_{k} t}$, while at Re$\,t \to -\infty$ the positive frequency
part is rescaled by the parameter $e^\theta$:  $\phi({\bf k})=f_{\bf k}\,
e^{-i\,\omega_{k} t}+ e^\theta \, f_{-\bf k}^*\, e^{i\,\omega_{k} t}$. The
parameter $\theta$ in the boundary condition and the parameter $T$ of the
contour (cf. figure 3) are Legendre-conjugate of respectively the
multiplicity $n_i$ and the total energy $E$ of initial particles. Namely, if
$i\, S$ is the classical action on the whole contour (thus $S$ is defined in
the way that it is real in the Euclidean space), one finds\cite{rst2}
\beq
n_i=2\, {{\partial S} \over {\partial \theta}}~,~~~~~
E=2\, {{\partial S} \over {\partial T}}~.
\label{entt}
\eeq
Furthermore the ``holy grail" function, entering the WKB estimate of the
total cross section as $\sigma^{tot} \sim \exp \left(-g^{-2}\, F(\epsilon,
\nu)\right)$ is given by the Legendre transform of the action:
\beq
{1 \over {g^2}}\, F(\epsilon, \nu)= 2\,S-ET-n_i \theta~~.
\label{fenu}
\eeq

Quite naturally, the formulated classical boundary value problem is not
easily solvable, and a sufficiently good approximation to the solution is
known only in a few models[53 - 55]. In particular the model,
considered in \cite{sr}, describes one scalar field
in (1+1) dimensions with the potential
\beq
  V(\phi) =    {m^2\over 2}\phi^2 -
 {m^2v^2 \over 2}\exp\left[2\lambda\left({\phi\over v}-1\right)\right]~,
  \label{potential}
\eeq
where $v$ and $\lambda$ are dimensionless constants, which both are assumed
to be large. The parameter $1/v$ is the small coupling constant of the
perturbation theory in the vacuum $\phi=0$. The negative sign of the
interaction term implies that the energy is unbounded from below at large
$\phi$, thus the vacuum $\phi=0$ is metastable, and is separated from the
decreasing part of the potential by a barrier located at $\phi \approx v$,
provided that $\lambda \gg 1$. Beyond the maximum the potential rapidly goes
down, so that the potential essentially is a quadratic well with a
``cliff"\cite{sr}. The metastability of the perturbative vacuum at $\phi=0$
does not show up in calculations of the scattering amplitudes to any finite
order of the perturbation theory, and it only arises through a
non-perturbative effect: unitary ``shadow" from the false vacuum decay,
which makes this contribution analogous to instanton-induced scattering
amplitudes in a Yang-Mills theory\cite{mv2,hsu}. The analog of the sphaleron
energy is the height of the barrier separating two phases: $E_{\rm Sp} =
{\rm const}\cdot m v^2$.

At large $\lambda$ the potential (\ref{potential}) contains a sharp
matching of the quadratic part (free field) and  a steep exponential
``cliff", which enables\cite{sr} to solve the boundary value problem in the
leading order in $1/\lambda$ and also to clarify the contribution of
multi-instanton (multi-bounce) configurations. It has been found\cite{sr}
that the multi-instanton configurations in this model are still not
important when the one-instanton contribution becomes large. As a result the
``holy grail" function, as shown in figure 4, reaches zero at finite energy,
which energy increases when the semiclassical parameter of the initial state
multiplicity $\nu=n_i\,v^2$ decreases. In figure 4 is also shown the
behavior corresponding to the periodic instanton, which maximizes over
$n_i$ the rate of tunneling through the barrier in the processes $n_i \to
n_f$ at given energy $E$ \cite{krt}.

\begin{figure}
\epsfbox[-97 35 143 156]{figure4.ps}
\caption{The ``holy grail"
function normalized to $F(0,\nu)=1$ in the (1+1) dimensional model[54]
with exponential interaction. }
\end{figure}

\subsubsection{Prospects for QCD hard processes.} It has been
argued[57 - 60] that a manifestation of instanton-induced
scattering in a weak coupling regime can be observed in hard processes in
QCD. The suggestion is to search for final states in hadron collisions,
which contain a large number of minijets, each with a typical invariant mass
$\mu$, such that $\alpha_s(\mu)$ is sufficiently small, e.g. $\mu
\approx 4\,$GeV, so that $\alpha_s(\mu) \approx 0.25$. An instanton-induced
process should involve production of typically $n_j \approx
4\pi/\alpha_s(\mu) \approx 50$ such jets, which requires energy in a parton
- parton collision of somewhat higher than $n_j \, \mu \approx 200\,$GeV.
The prospects of observing the instanton induced hard processes in QCD are
certainly more phenomenologically attractive, since, unlike the electroweak
case, the energy range is hopefully within the reach of LHC and also the
cross section can be of a more encouraging magnitude, even if it is
suppressed by an exponential factor, like $\exp(-2\pi/\alpha_s(\mu)) \sim
10^{-11}$ as suggested by the ``premature unitarization" models.
However the reality of observing these possible non-perturbative hard
processes in QCD is still under discussion.

\section{Multi-particle production in topologically trivial sector}

The growth of the rate of the instanton-induced
processes is associated with production of multiboson final states until at
high multiplicity $n_f \sim 1/g^2$ the final state becomes not tractable
perturbatively. A similar problem in fact arises\cite{cornwall,goldberg} at
those high multiplicities in processes, which do not require contribution of
field configurations with non-trivial topology, and thus are allowed in
perturbation theory. This is related to the well known factorial growth of
coefficients in the perturbation theory series\cite{dyson}. Namely in the
perturbation theory the total cross section for production of $n$ bosons
interacting with a weak coupling $g$ is given, modulo the phase space
suppression at finite energy, by
\beq
\sigma_n \sim n!\, (g^2)^n~.
\label{sigman}
\eeq
At small $n$, naturally, the cross section is decreasing with multiplicity.
However at $n \sim 1/g^2$ the growth of $n!$ becomes faster than the
decrease of $(g^2)^n$ and the behavior (\ref{sigman}) would imply that the
cross section starts to grow with multiplicity. Therefore the question: ``If
there is enough energy to produce $\gg 1/\alpha_W$ $W,\,Z,\,H$ bosons, will
they be actually produced with non-negligible cross section?" does not seem
to be entirely paradoxical or idle in view of eq.(\ref{sigman}).
The difficulty of answering this question is in that the lowest order
equation (\ref{sigman}) becomes inapplicable already at $n \sim 1/g$, since
the loop corrections to $\sigma_n$ are governed by the parameter $n^2 \,
g^2$. The latter can be seen from the number of rescatterings between the
final particles: $O(n^2)$, each having strength $g^2$.
In what follows we will discuss
several steps that have been attempted toward answering the above question.
It also may well be that a solution of the multiboson problem without the
topological complications will provide an insight into the problem of (B+L)
violation in high-energy collisions.

\subsection{Multiboson amplitudes in $\lambda \, \phi^4$ theories.}

One of simplest models, where the development of non-perturbative dynamics
in multiboson amplitudes can be studied, is that describing one real scalar
field with the $\lambda \, \phi^4$ interaction, whose potential is given by
\beq
V(\phi)={{m^2} \over 2 }\, \phi^2 +{\lambda \over 4} \, \phi^4~.
\label{phi4}
\eeq
If $m^2$ is positive the field has one vacuum state at $\langle 0 | \phi | 0
\rangle =0$ and the symmetry under sign reflection: $\phi \to -\phi$ is
unbroken, while at negative $m^2$ there are two degenerate vacua
$\langle 0 | \phi | 0
\rangle = \pm v$ with $v=|m|/\sqrt{\lambda}$, which situation describes the
spontaneous symmetry breaking (SSB).

\subsubsection{Multiboson amplitudes at zero energy and momentum.}
The simplest problem concerning multiboson amplitudes is, perhaps, that of
calculating connected $n$-boson off-shell scattering amplitudes $A_n$, in
which all the external particles have zero energy and
momentum\cite{golv,mv4}.
The amplitude $A_n$ can be written in terms of the connected part of the
Euclidean-space correlator:
\beq
{{\int \left ( \int \sigma(x)\, d^d x \right )^n \, \exp (-S[\phi])\, {\cal
D}\phi} \over {\int \exp (-S[\phi])\, {\cal D}\phi}}~,
\label{defin}
\eeq
where $\sigma(x)$ is the deviation of the field $\phi$ from its vacuum mean
value $\sigma(x)=\phi (x) - \langle 0|\phi| 0 \rangle $, and the integral
$\int \sigma(x) d^d x$ is understood as the $p \to 0$ limit of the
Fourier transform $\int \sigma(x) e^{i p x} d^d x$.  Furthermore the
connected part of the correlator (\ref{defin}) is conveniently given by the
$n$-th logarithmic derivative of the generating functional $Z(j)= \int \exp
\bigl ( -S[\phi]+ \int j \sigma(x) \, d^d x \bigr ) \, {\cal D}\phi$ with a
constant source $j$:
\beq
A_n=\left . \left ( {d \over {dj}} \right )^n \ln Z(j) \right |_{j=0}~.
\label{azj}
\eeq
Introduction of a constant source is equivalent to replacing the original
potential $V(\phi)$ by $V(\phi)-j\, \sigma$. Furthermore, for a constant
source $\ln Z(j)$ is related to the energy of the vacuum $E(j)$ in the
presence of $j$: $\ln Z(j)=-VT \, E(j)$, where $VT$ is the normalization
space-time volume. Thus according to eq.(\ref{azj}) the asymptotic at large
$n$ behavior of the amplitudes $A_n$ is related to the position $j_c$ of the
nearest to $j=0$ singularity of $E(j)$ in the complex $j$ plane: $A_n \sim
n! j_c^{-n}$. At the classical level the position of the singularity is
determined by the value of $j$, at which two solutions of the equilibrium
equation $dV/d\phi = j$ coincide. For the potential (\ref{phi4}) this
happens at $j_c = \pm i\,\sqrt{4/27\lambda}\, m^3$, which determines the
asymptotic behavior\cite{gv,mv4} of the tree-level amplitudes $A_n$:
\beq
|A_n^{\rm tree}| \sim n! \left ( {27 \over 4} \, {\lambda \over |m|^6}
\right )^{n/2}~.
\label{antree}
\eeq

In a theory with unbroken symmetry the quantum loops modify $E(j)$ according
to the Coleman-Weinberg potential\cite{cw} thus shifting and modifying the
singularity in the $j$ plane. However these corrections neither eliminate
the singularity nor bring it to $j=0$. The shift of the position can be
absorbed in normalization of $\lambda$ and $m$, while the modification of
the type of the singularity only affects sub-leading in $n$ factors, so that
the leading behavior in eq.(\ref{antree}) is not modified by quantum
corrections in a theory with unbroken symmetry.

The situation with quantum effects in a theory with SSB is drastically
different: non-perturbatively the point $j=0$ is in fact a branch point
of the vacuum energy $E(j)$ for either of the vacua. Indeed,
if, for definiteness, one choses to consider the amplitudes $A_n$
in the `left' vacuum: $\langle 0 | \phi | 0
\rangle = - v$ with $v=|m|/\sqrt{\lambda}$, and follows the dependence of
its energy on $j$, one finds that this state is stable for real $j < 0$ and
is metastable at arbitrarily small positive $j$. Thus at $j > 0$ the
energy $E(j)$ acquires an imaginary part given by the decay rate of the
metastable vacuum.  In this situation the Taylor expansion of $E(j)$ is
asymptotic and the coefficients are determined by the decay rate in the
presence of an infinitesimal positive source term. In this situation the
calculation\cite{vko} of the false vacuum decay rate
in the thin wall approximation is applicable exactly. Thus one can readily
find the exact non-perturbative asymptotic behavior of the amplitudes $A_n$
at large $n$ in a theory in $d$ space-time dimensions\cite{mv4}:
\beq
A_n \sim \left ( {{n\,d} \over {d-1}} \right )! \left ( {{c_d \, \lambda}
\over {|m|^6}} \right )^{n \over 2} \left ( {\lambda \over {|m|^{4-d}}}
\right )^{n \over {d-1}}
\label{anex}
\eeq
with
$$ c_d=\left [ {3^d \over {2^{2d-1}}} \, {{\Gamma (d/2)} \over {d^{d-1} \,
\pi^{d/2} }} \right ] ^{2 \over {d-1}}~.$$
The factorial behavior in eq.(\ref{anex}), if interpreted in terms of loop
graphs in perturbation theory, corresponds to contribution of graphs with
$n/(d-1)$ loops.

The considered off-shell amplitudes $A_n$ are not physical. However one can
draw from the described exercise at least two, possibly important,
theoretical conclusions about multiboson amplitudes: \\
$\bullet$ the $n!$ behavior suggested by the tree-level analysis
is not necessarily eliminated and may even be enhanced in the exact result,
and \\
$\bullet$ the large $n$ behavior of multiboson amplitudes does not have to
be universal and may in fact be very sensitive to details of the theory.

\subsubsection{Production of on-shell multiparticle states at and above
threshold. Tree graphs.}

More physical, than the previously discussed off-shell amplitudes, are the
amplitudes of processes, where $n$ on-shell bosons are produced by a highly
virtual field $\phi$ ($1 \to n$ process): $a_n=\langle n | \phi(0) | 0
\rangle$. (These e.g.  can be related to the reaction $e^+ e^- \to n\, H$.)
As will be explained, it turns out that one can explicitly find the sum of
all tree graphs and all one-loop graphs for these amplitudes at any $n$,
provided that the final bosons are exactly at rest in the c.m. system.  Also
the summation of two- and higher- loop graphs is in principle possible for
this kinematical arrangement, however a calculation with a finite number of
loops is inevitably plagued by the breakdown of the perturbation theory at
large $n$.  Thus far three methods have been used in calculation of the
threshold amplitudes of the $1 \to n$ processes: the Landau WKB method,
the recursion equations, and the functional technique.

Landau WKB method\cite{landau,ll}
is used in Quantum Mechanics for calculating transition
matrix elements between strongly different levels. (For a field theory
derivation of this technique see \cite{ip}.) In the tree
graphs for the threshold $1 \to n$ amplitudes all the external and
internal lines carry no spatial momentum. Thus the problem is reduced to
dynamics of only one mode of the field with spatial momentum ${\bf p}=0$,
i.e. to a Quantum Mechanical problem. This approach yields the
result\cite{mv5} for the sum of the tree graphs at the threshold with
accuracy up to terms $O(1/n^2)$ at large $n$.  (Application of the Landau
WKB technique in the problem of multiboson amplitudes is also discussed in
\cite{porrati2,cortik,khlebnikov,dp2}.)

{Recursion equations}\cite{mv4} for the amplitudes $a(n)$ arise from
inspecting the construction of Feynman graphs. For the simplest case of tree
graphs in $\lambda \phi^4$ theory the algebraic
form of the equations is
\begin{eqnarray}
&&(n^2-1)\,{a(n) \over n!} =
\nonumber \\
&&\lambda \,\sum_{odd~n_1,\,n_2} {a(n_1) \over n_1!}\,{a(n_2)
\over n_2!}\,{a(n-n_1-n_2) \over {(n-n_1-n_2)!}}~,
\label{recur}
\end{eqnarray}
where the sum runs over odd $n_1$ and $n_2$ as well as $n$ is
odd, since due to the unbroken sign reflection symmetry the parity of the
number of particles is conserved. Also the mass $m$ in eq.(\ref{recur}) is
set to one, since it can be restored in the final result from dimensional
counting.  The solution to  the equations (\ref{recur}) reads as\cite{mv4}
\beq
a(n)=n!\,  ( \lambda /8 m^2  )^{{n-1} \over
2} ~,
\label{sol1}
\eeq
which can be found by applying the regular method of generating
functions\cite{akp1}. For the theory with SSB the recursion equations are
modified by the presence of cubic vertices. The result for the amplitudes in
the theory with SSB is\cite{akp1}
\beq
a(n)=-n!\, ( 2\, v)^{1-n}
\label{sol2}
\eeq

The recursion method can be extended to other theories\cite{akp3}
as well as to
loop graphs\cite{mv6,akp4} and to an analysis of higher loops\cite{akp6}.
However a more convenient method for
further analysis is the one suggested by Brown\cite{brown} and is based on
a functional technique. Before proceeding to discussing this method and its
further applications we report on estimates of the tree amplitudes above the
threshold and thus of the total probability of the processes $1 \to n$ at
a high energy $E$.

\subsubsection{Lower bound for cross section at the tree level.}
The tree graphs for the processes $1 \to n$ in a $\lambda \phi^4$ theory all
have the same sign\cite{cornwall,goldberg}. The decrease of the amplitude
above the threshold is thus determined by the increasing virtuality of the
propagators in those graphs, which depends on the kinematics of the final
state. One can thus find a lower bound on the tree amplitudes above the
threshold in a restricted part of the final-particle phase
space[79 - 81], which gives a lower bound on the total probability
of the process. In particular, if the kinematical restriction is
chosen\cite{mv7,mv8} as the condition that the c.m. energy of each
individual particle in the final state does not exceed $\omega$, then in
this region of the phase space the tree amplitude $A(1 \to n)$ is larger
than the threshold amplitude $a(n)$ in which the physical mass $M$ of the
scalar boson is replaced by ${9 \over 8} \omega $ in the theory without SSB
(in which case $M=m$) and by ${4 \over 3} \omega $ in the theory with SSB
(where $M=\sqrt{2}\,|m|$). The cut off energy $\omega$ is then optimized for
each value of the total energy $E$ and multiplicity $n$ in order to find the
largest lower bound on the total probability
\beq
\sigma_n = \int \left | A(1 \to n) \right |^2 \, d\tau_n~,
\eeq
which is given by the integral over the $n$ particle phase space $\tau_n$.
As a result the lower bound on $\sigma_n$ is found\cite{mv8}
in the scaling form
\beq
\sigma_n > \exp \left [ {{4\pi^2 c} \over \lambda} f(\epsilon,\nu) \right ]
{}~,
\label{scal}
\eeq
where $\nu=n \,M/E$ is the ratio of the multiplicity $n$ to its maximal
possible value $E/M$, $\epsilon=E/E_0$ with $E_0$ being an analog of the
`sphaleron' energy: $E_0=4\pi^2 \,c \, M/\lambda$, and the constant $c$ in
these formulas is $c=9~(c=8/3)$ for a theory without (with) SSB.
The calculated\cite{mv8} behavior of the function $f(\epsilon,\nu)$ is shown
in figure 5, which thus illustrates and quantifies the interplay between the
$n!$ and the power of small coupling constant, discussed in connection with
eq.(\ref{sigman}).  The function $f(\epsilon,\nu)$ displays a normal
perturbative maximum at zero multiplicity. However, as energy grows, and
production of high multiplicity states becomes unsuppressed kinematically,
this function develops a second maximum, which at larger energies eventually
crosses zero with an apparent violation of unitarity.

It is interesting to notice that the kinematical suppression of
multiparticle final states is quite essential and shifts the energy, at
which the tree graphs violate unitarity significantly higher than one would
guess from a simple estimate $E_{crit} \approx 4\pi\,M/\lambda$. If applied
to a multi-Higgs production in the Standard Model the lower bound
(\ref{scal}) breaks unitarity at $E_{crit} \approx 15.5 \, (32\pi^2
M_H/\lambda) \approx 1000 \, {\rm GeV} (200 {\rm GeV}/M_H)$.

It should be also mentioned that it is quite likely that the scaling
behavior (\ref{scal}) at a given large multiplicity $n$ also holds for the
actual cross section, which point is recently strongly emphasized in
\cite{lrst}. The function $f(\epsilon, \nu)$ can thus be called differential
in $n$ ``holy grail" function.

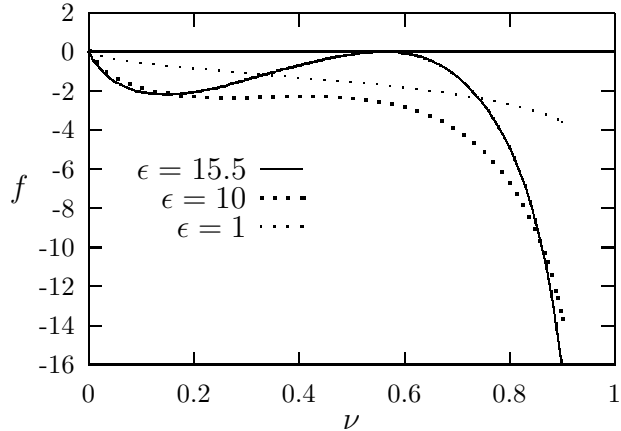
\begin{figure} \begin{center}
\setlength{\unitlength}{0.240900pt}
\ifx\plotpoint\undefined\newsavebox{\plotpoint}\fi
\sbox{\plotpoint}{\rule[-0.200pt]{0.400pt}{0.400pt}}%
\begin{picture}(1000,689)(95,10)
\font\gnuplot=cmr10 at 10pt
\gnuplot
\sbox{\plotpoint}{\rule[-0.200pt]{0.400pt}{0.400pt}}%
\put(220.0,605.0){\rule[-0.200pt]{198.983pt}{0.400pt}}
\put(220.0,113.0){\rule[-0.200pt]{0.400pt}{133.218pt}}
\put(220.0,113.0){\rule[-0.200pt]{4.818pt}{0.400pt}}
\put(198,113){\makebox(0,0)[r]{-16}}
\put(1026.0,113.0){\rule[-0.200pt]{4.818pt}{0.400pt}}
\put(220.0,174.0){\rule[-0.200pt]{4.818pt}{0.400pt}}
\put(198,174){\makebox(0,0)[r]{-14}}
\put(1026.0,174.0){\rule[-0.200pt]{4.818pt}{0.400pt}}
\put(220.0,236.0){\rule[-0.200pt]{4.818pt}{0.400pt}}
\put(198,236){\makebox(0,0)[r]{-12}}
\put(1026.0,236.0){\rule[-0.200pt]{4.818pt}{0.400pt}}
\put(220.0,297.0){\rule[-0.200pt]{4.818pt}{0.400pt}}
\put(198,297){\makebox(0,0)[r]{-10}}
\put(1026.0,297.0){\rule[-0.200pt]{4.818pt}{0.400pt}}
\put(220.0,359.0){\rule[-0.200pt]{4.818pt}{0.400pt}}
\put(198,359){\makebox(0,0)[r]{-8}}
\put(1026.0,359.0){\rule[-0.200pt]{4.818pt}{0.400pt}}
\put(220.0,420.0){\rule[-0.200pt]{4.818pt}{0.400pt}}
\put(198,420){\makebox(0,0)[r]{-6}}
\put(1026.0,420.0){\rule[-0.200pt]{4.818pt}{0.400pt}}
\put(220.0,482.0){\rule[-0.200pt]{4.818pt}{0.400pt}}
\put(198,482){\makebox(0,0)[r]{-4}}
\put(1026.0,482.0){\rule[-0.200pt]{4.818pt}{0.400pt}}
\put(220.0,543.0){\rule[-0.200pt]{4.818pt}{0.400pt}}
\put(198,543){\makebox(0,0)[r]{-2}}
\put(1026.0,543.0){\rule[-0.200pt]{4.818pt}{0.400pt}}
\put(220.0,605.0){\rule[-0.200pt]{4.818pt}{0.400pt}}
\put(198,605){\makebox(0,0)[r]{0}}
\put(1026.0,605.0){\rule[-0.200pt]{4.818pt}{0.400pt}}
\put(220.0,666.0){\rule[-0.200pt]{4.818pt}{0.400pt}}
\put(198,666){\makebox(0,0)[r]{2}}
\put(1026.0,666.0){\rule[-0.200pt]{4.818pt}{0.400pt}}
\put(220.0,113.0){\rule[-0.200pt]{0.400pt}{4.818pt}}
\put(220,68){\makebox(0,0){0}}
\put(220.0,646.0){\rule[-0.200pt]{0.400pt}{4.818pt}}
\put(385.0,113.0){\rule[-0.200pt]{0.400pt}{4.818pt}}
\put(385,68){\makebox(0,0){0.2}}
\put(385.0,646.0){\rule[-0.200pt]{0.400pt}{4.818pt}}
\put(550.0,113.0){\rule[-0.200pt]{0.400pt}{4.818pt}}
\put(550,68){\makebox(0,0){0.4}}
\put(550.0,646.0){\rule[-0.200pt]{0.400pt}{4.818pt}}
\put(716.0,113.0){\rule[-0.200pt]{0.400pt}{4.818pt}}
\put(716,68){\makebox(0,0){0.6}}
\put(716.0,646.0){\rule[-0.200pt]{0.400pt}{4.818pt}}
\put(881.0,113.0){\rule[-0.200pt]{0.400pt}{4.818pt}}
\put(881,68){\makebox(0,0){0.8}}
\put(881.0,646.0){\rule[-0.200pt]{0.400pt}{4.818pt}}
\put(1046.0,113.0){\rule[-0.200pt]{0.400pt}{4.818pt}}
\put(1046,68){\makebox(0,0){1}}
\put(1046.0,646.0){\rule[-0.200pt]{0.400pt}{4.818pt}}
\put(220.0,113.0){\rule[-0.200pt]{198.983pt}{0.400pt}}
\put(1046.0,113.0){\rule[-0.200pt]{0.400pt}{133.218pt}}
\put(220.0,666.0){\rule[-0.200pt]{198.983pt}{0.400pt}}
\put(111,389){\makebox(0,0){$f$}}
\put(633,23){\makebox(0,0){$\nu$}}
\put(220.0,113.0){\rule[-0.200pt]{0.400pt}{133.218pt}}
\put(468,420){\makebox(0,0)[r]{$\epsilon=15.5$}}
\put(490.0,420.0){\rule[-0.200pt]{15.899pt}{0.400pt}}
\put(220,605){\usebox{\plotpoint}}
\multiput(220.59,601.68)(0.488,-0.890){13}{\rule{0.117pt}{0.800pt}}
\multiput(219.17,603.34)(8.000,-12.340){2}{\rule{0.400pt}{0.400pt}}
\multiput(228.59,588.56)(0.489,-0.611){15}{\rule{0.118pt}{0.589pt}}
\multiput(227.17,589.78)(9.000,-9.778){2}{\rule{0.400pt}{0.294pt}}
\multiput(237.59,577.72)(0.488,-0.560){13}{\rule{0.117pt}{0.550pt}}
\multiput(236.17,578.86)(8.000,-7.858){2}{\rule{0.400pt}{0.275pt}}
\multiput(245.00,569.93)(0.494,-0.488){13}{\rule{0.500pt}{0.117pt}}
\multiput(245.00,570.17)(6.962,-8.000){2}{\rule{0.250pt}{0.400pt}}
\multiput(253.00,561.93)(0.671,-0.482){9}{\rule{0.633pt}{0.116pt}}
\multiput(253.00,562.17)(6.685,-6.000){2}{\rule{0.317pt}{0.400pt}}
\multiput(261.00,555.94)(1.212,-0.468){5}{\rule{1.000pt}{0.113pt}}
\multiput(261.00,556.17)(6.924,-4.000){2}{\rule{0.500pt}{0.400pt}}
\multiput(270.00,551.94)(1.066,-0.468){5}{\rule{0.900pt}{0.113pt}}
\multiput(270.00,552.17)(6.132,-4.000){2}{\rule{0.450pt}{0.400pt}}
\multiput(278.00,547.95)(1.579,-0.447){3}{\rule{1.167pt}{0.108pt}}
\multiput(278.00,548.17)(5.579,-3.000){2}{\rule{0.583pt}{0.400pt}}
\multiput(286.00,544.95)(1.579,-0.447){3}{\rule{1.167pt}{0.108pt}}
\multiput(286.00,545.17)(5.579,-3.000){2}{\rule{0.583pt}{0.400pt}}
\put(294,541.17){\rule{1.900pt}{0.400pt}}
\multiput(294.00,542.17)(5.056,-2.000){2}{\rule{0.950pt}{0.400pt}}
\put(303,539.67){\rule{1.927pt}{0.400pt}}
\multiput(303.00,540.17)(4.000,-1.000){2}{\rule{0.964pt}{0.400pt}}
\put(311,538.67){\rule{1.927pt}{0.400pt}}
\multiput(311.00,539.17)(4.000,-1.000){2}{\rule{0.964pt}{0.400pt}}
\put(319,537.67){\rule{1.927pt}{0.400pt}}
\multiput(319.00,538.17)(4.000,-1.000){2}{\rule{0.964pt}{0.400pt}}
\put(336,536.67){\rule{1.927pt}{0.400pt}}
\multiput(336.00,537.17)(4.000,-1.000){2}{\rule{0.964pt}{0.400pt}}
\put(344,536.67){\rule{1.927pt}{0.400pt}}
\multiput(344.00,536.17)(4.000,1.000){2}{\rule{0.964pt}{0.400pt}}
\put(327.0,538.0){\rule[-0.200pt]{2.168pt}{0.400pt}}
\put(360,537.67){\rule{2.168pt}{0.400pt}}
\multiput(360.00,537.17)(4.500,1.000){2}{\rule{1.084pt}{0.400pt}}
\put(369,538.67){\rule{1.927pt}{0.400pt}}
\multiput(369.00,538.17)(4.000,1.000){2}{\rule{0.964pt}{0.400pt}}
\put(377,539.67){\rule{1.927pt}{0.400pt}}
\multiput(377.00,539.17)(4.000,1.000){2}{\rule{0.964pt}{0.400pt}}
\put(385,540.67){\rule{1.927pt}{0.400pt}}
\multiput(385.00,540.17)(4.000,1.000){2}{\rule{0.964pt}{0.400pt}}
\put(393,542.17){\rule{1.900pt}{0.400pt}}
\multiput(393.00,541.17)(5.056,2.000){2}{\rule{0.950pt}{0.400pt}}
\put(402,544.17){\rule{1.700pt}{0.400pt}}
\multiput(402.00,543.17)(4.472,2.000){2}{\rule{0.850pt}{0.400pt}}
\put(410,545.67){\rule{1.927pt}{0.400pt}}
\multiput(410.00,545.17)(4.000,1.000){2}{\rule{0.964pt}{0.400pt}}
\put(418,547.17){\rule{1.900pt}{0.400pt}}
\multiput(418.00,546.17)(5.056,2.000){2}{\rule{0.950pt}{0.400pt}}
\put(427,549.17){\rule{1.700pt}{0.400pt}}
\multiput(427.00,548.17)(4.472,2.000){2}{\rule{0.850pt}{0.400pt}}
\multiput(435.00,551.61)(1.579,0.447){3}{\rule{1.167pt}{0.108pt}}
\multiput(435.00,550.17)(5.579,3.000){2}{\rule{0.583pt}{0.400pt}}
\put(443,554.17){\rule{1.700pt}{0.400pt}}
\multiput(443.00,553.17)(4.472,2.000){2}{\rule{0.850pt}{0.400pt}}
\put(451,556.17){\rule{1.900pt}{0.400pt}}
\multiput(451.00,555.17)(5.056,2.000){2}{\rule{0.950pt}{0.400pt}}
\multiput(460.00,558.61)(1.579,0.447){3}{\rule{1.167pt}{0.108pt}}
\multiput(460.00,557.17)(5.579,3.000){2}{\rule{0.583pt}{0.400pt}}
\put(468,561.17){\rule{1.700pt}{0.400pt}}
\multiput(468.00,560.17)(4.472,2.000){2}{\rule{0.850pt}{0.400pt}}
\put(476,563.17){\rule{1.700pt}{0.400pt}}
\multiput(476.00,562.17)(4.472,2.000){2}{\rule{0.850pt}{0.400pt}}
\multiput(484.00,565.61)(1.802,0.447){3}{\rule{1.300pt}{0.108pt}}
\multiput(484.00,564.17)(6.302,3.000){2}{\rule{0.650pt}{0.400pt}}
\put(493,568.17){\rule{1.700pt}{0.400pt}}
\multiput(493.00,567.17)(4.472,2.000){2}{\rule{0.850pt}{0.400pt}}
\put(501,570.17){\rule{1.700pt}{0.400pt}}
\multiput(501.00,569.17)(4.472,2.000){2}{\rule{0.850pt}{0.400pt}}
\multiput(509.00,572.61)(1.579,0.447){3}{\rule{1.167pt}{0.108pt}}
\multiput(509.00,571.17)(5.579,3.000){2}{\rule{0.583pt}{0.400pt}}
\put(517,575.17){\rule{1.900pt}{0.400pt}}
\multiput(517.00,574.17)(5.056,2.000){2}{\rule{0.950pt}{0.400pt}}
\put(526,577.17){\rule{1.700pt}{0.400pt}}
\multiput(526.00,576.17)(4.472,2.000){2}{\rule{0.850pt}{0.400pt}}
\put(534,579.17){\rule{1.700pt}{0.400pt}}
\multiput(534.00,578.17)(4.472,2.000){2}{\rule{0.850pt}{0.400pt}}
\put(542,581.17){\rule{1.700pt}{0.400pt}}
\multiput(542.00,580.17)(4.472,2.000){2}{\rule{0.850pt}{0.400pt}}
\put(550,583.17){\rule{1.900pt}{0.400pt}}
\multiput(550.00,582.17)(5.056,2.000){2}{\rule{0.950pt}{0.400pt}}
\put(559,585.17){\rule{1.700pt}{0.400pt}}
\multiput(559.00,584.17)(4.472,2.000){2}{\rule{0.850pt}{0.400pt}}
\put(567,587.17){\rule{1.700pt}{0.400pt}}
\multiput(567.00,586.17)(4.472,2.000){2}{\rule{0.850pt}{0.400pt}}
\put(575,589.17){\rule{1.700pt}{0.400pt}}
\multiput(575.00,588.17)(4.472,2.000){2}{\rule{0.850pt}{0.400pt}}
\put(583,591.17){\rule{1.900pt}{0.400pt}}
\multiput(583.00,590.17)(5.056,2.000){2}{\rule{0.950pt}{0.400pt}}
\put(592,593.17){\rule{1.700pt}{0.400pt}}
\multiput(592.00,592.17)(4.472,2.000){2}{\rule{0.850pt}{0.400pt}}
\put(600,594.67){\rule{1.927pt}{0.400pt}}
\multiput(600.00,594.17)(4.000,1.000){2}{\rule{0.964pt}{0.400pt}}
\put(608,596.17){\rule{1.700pt}{0.400pt}}
\multiput(608.00,595.17)(4.472,2.000){2}{\rule{0.850pt}{0.400pt}}
\put(616,597.67){\rule{2.168pt}{0.400pt}}
\multiput(616.00,597.17)(4.500,1.000){2}{\rule{1.084pt}{0.400pt}}
\put(625,598.67){\rule{1.927pt}{0.400pt}}
\multiput(625.00,598.17)(4.000,1.000){2}{\rule{0.964pt}{0.400pt}}
\put(633,600.17){\rule{1.700pt}{0.400pt}}
\multiput(633.00,599.17)(4.472,2.000){2}{\rule{0.850pt}{0.400pt}}
\put(641,601.67){\rule{2.168pt}{0.400pt}}
\multiput(641.00,601.17)(4.500,1.000){2}{\rule{1.084pt}{0.400pt}}
\put(352.0,538.0){\rule[-0.200pt]{1.927pt}{0.400pt}}
\put(658,602.67){\rule{1.927pt}{0.400pt}}
\multiput(658.00,602.17)(4.000,1.000){2}{\rule{0.964pt}{0.400pt}}
\put(650.0,603.0){\rule[-0.200pt]{1.927pt}{0.400pt}}
\put(674,603.67){\rule{2.168pt}{0.400pt}}
\multiput(674.00,603.17)(4.500,1.000){2}{\rule{1.084pt}{0.400pt}}
\put(666.0,604.0){\rule[-0.200pt]{1.927pt}{0.400pt}}
\put(691,603.67){\rule{1.927pt}{0.400pt}}
\multiput(691.00,604.17)(4.000,-1.000){2}{\rule{0.964pt}{0.400pt}}
\put(683.0,605.0){\rule[-0.200pt]{1.927pt}{0.400pt}}
\put(707,602.17){\rule{1.900pt}{0.400pt}}
\multiput(707.00,603.17)(5.056,-2.000){2}{\rule{0.950pt}{0.400pt}}
\put(716,600.67){\rule{1.927pt}{0.400pt}}
\multiput(716.00,601.17)(4.000,-1.000){2}{\rule{0.964pt}{0.400pt}}
\put(724,599.17){\rule{1.700pt}{0.400pt}}
\multiput(724.00,600.17)(4.472,-2.000){2}{\rule{0.850pt}{0.400pt}}
\put(732,597.17){\rule{1.700pt}{0.400pt}}
\multiput(732.00,598.17)(4.472,-2.000){2}{\rule{0.850pt}{0.400pt}}
\multiput(740.00,595.95)(1.802,-0.447){3}{\rule{1.300pt}{0.108pt}}
\multiput(740.00,596.17)(6.302,-3.000){2}{\rule{0.650pt}{0.400pt}}
\multiput(749.00,592.95)(1.579,-0.447){3}{\rule{1.167pt}{0.108pt}}
\multiput(749.00,593.17)(5.579,-3.000){2}{\rule{0.583pt}{0.400pt}}
\multiput(757.00,589.94)(1.066,-0.468){5}{\rule{0.900pt}{0.113pt}}
\multiput(757.00,590.17)(6.132,-4.000){2}{\rule{0.450pt}{0.400pt}}
\multiput(765.00,585.94)(1.066,-0.468){5}{\rule{0.900pt}{0.113pt}}
\multiput(765.00,586.17)(6.132,-4.000){2}{\rule{0.450pt}{0.400pt}}
\multiput(773.00,581.93)(0.933,-0.477){7}{\rule{0.820pt}{0.115pt}}
\multiput(773.00,582.17)(7.298,-5.000){2}{\rule{0.410pt}{0.400pt}}
\multiput(782.00,576.93)(0.671,-0.482){9}{\rule{0.633pt}{0.116pt}}
\multiput(782.00,577.17)(6.685,-6.000){2}{\rule{0.317pt}{0.400pt}}
\multiput(790.00,570.93)(0.671,-0.482){9}{\rule{0.633pt}{0.116pt}}
\multiput(790.00,571.17)(6.685,-6.000){2}{\rule{0.317pt}{0.400pt}}
\multiput(798.00,564.93)(0.569,-0.485){11}{\rule{0.557pt}{0.117pt}}
\multiput(798.00,565.17)(6.844,-7.000){2}{\rule{0.279pt}{0.400pt}}
\multiput(806.00,557.93)(0.560,-0.488){13}{\rule{0.550pt}{0.117pt}}
\multiput(806.00,558.17)(7.858,-8.000){2}{\rule{0.275pt}{0.400pt}}
\multiput(815.59,548.72)(0.488,-0.560){13}{\rule{0.117pt}{0.550pt}}
\multiput(814.17,549.86)(8.000,-7.858){2}{\rule{0.400pt}{0.275pt}}
\multiput(823.59,539.72)(0.488,-0.560){13}{\rule{0.117pt}{0.550pt}}
\multiput(822.17,540.86)(8.000,-7.858){2}{\rule{0.400pt}{0.275pt}}
\multiput(831.59,530.74)(0.489,-0.553){15}{\rule{0.118pt}{0.544pt}}
\multiput(830.17,531.87)(9.000,-8.870){2}{\rule{0.400pt}{0.272pt}}
\multiput(840.59,520.30)(0.488,-0.692){13}{\rule{0.117pt}{0.650pt}}
\multiput(839.17,521.65)(8.000,-9.651){2}{\rule{0.400pt}{0.325pt}}
\multiput(848.59,508.89)(0.488,-0.824){13}{\rule{0.117pt}{0.750pt}}
\multiput(847.17,510.44)(8.000,-11.443){2}{\rule{0.400pt}{0.375pt}}
\multiput(856.59,495.89)(0.488,-0.824){13}{\rule{0.117pt}{0.750pt}}
\multiput(855.17,497.44)(8.000,-11.443){2}{\rule{0.400pt}{0.375pt}}
\multiput(864.59,482.82)(0.489,-0.844){15}{\rule{0.118pt}{0.767pt}}
\multiput(863.17,484.41)(9.000,-13.409){2}{\rule{0.400pt}{0.383pt}}
\multiput(873.59,467.26)(0.488,-1.022){13}{\rule{0.117pt}{0.900pt}}
\multiput(872.17,469.13)(8.000,-14.132){2}{\rule{0.400pt}{0.450pt}}
\multiput(881.59,450.85)(0.488,-1.154){13}{\rule{0.117pt}{1.000pt}}
\multiput(880.17,452.92)(8.000,-15.924){2}{\rule{0.400pt}{0.500pt}}
\multiput(889.59,432.23)(0.488,-1.352){13}{\rule{0.117pt}{1.150pt}}
\multiput(888.17,434.61)(8.000,-18.613){2}{\rule{0.400pt}{0.575pt}}
\multiput(897.59,411.53)(0.489,-1.252){15}{\rule{0.118pt}{1.078pt}}
\multiput(896.17,413.76)(9.000,-19.763){2}{\rule{0.400pt}{0.539pt}}
\multiput(906.59,388.19)(0.488,-1.682){13}{\rule{0.117pt}{1.400pt}}
\multiput(905.17,391.09)(8.000,-23.094){2}{\rule{0.400pt}{0.700pt}}
\multiput(914.59,361.57)(0.488,-1.880){13}{\rule{0.117pt}{1.550pt}}
\multiput(913.17,364.78)(8.000,-25.783){2}{\rule{0.400pt}{0.775pt}}
\multiput(922.59,331.74)(0.488,-2.145){13}{\rule{0.117pt}{1.750pt}}
\multiput(921.17,335.37)(8.000,-29.368){2}{\rule{0.400pt}{0.875pt}}
\multiput(930.59,298.57)(0.489,-2.184){15}{\rule{0.118pt}{1.789pt}}
\multiput(929.17,302.29)(9.000,-34.287){2}{\rule{0.400pt}{0.894pt}}
\multiput(939.59,258.45)(0.488,-2.871){13}{\rule{0.117pt}{2.300pt}}
\multiput(938.17,263.23)(8.000,-39.226){2}{\rule{0.400pt}{1.150pt}}
\multiput(947.59,213.00)(0.488,-3.333){13}{\rule{0.117pt}{2.650pt}}
\multiput(946.17,218.50)(8.000,-45.500){2}{\rule{0.400pt}{1.325pt}}
\multiput(955.59,160.34)(0.488,-3.862){13}{\rule{0.117pt}{3.050pt}}
\multiput(954.17,166.67)(8.000,-52.670){2}{\rule{0.400pt}{1.525pt}}
\put(699.0,604.0){\rule[-0.200pt]{1.927pt}{0.400pt}}
\sbox{\plotpoint}{\rule[-0.500pt]{1.000pt}{1.000pt}}%
\put(468,375){\makebox(0,0)[r]{$\epsilon=10$}}
\multiput(490,375)(20.756,0.000){4}{\usebox{\plotpoint}}
\put(556,375){\usebox{\plotpoint}}
\put(220,605){\usebox{\plotpoint}}
\put(220.00,605.00){\usebox{\plotpoint}}
\put(231.67,587.92){\usebox{\plotpoint}}
\multiput(237,582)(14.676,-14.676){0}{\usebox{\plotpoint}}
\put(246.18,573.11){\usebox{\plotpoint}}
\multiput(253,568)(17.601,-11.000){0}{\usebox{\plotpoint}}
\put(263.45,561.64){\usebox{\plotpoint}}
\multiput(270,558)(19.434,-7.288){0}{\usebox{\plotpoint}}
\put(282.42,553.34){\usebox{\plotpoint}}
\multiput(286,552)(19.434,-7.288){0}{\usebox{\plotpoint}}
\put(302.18,547.18){\usebox{\plotpoint}}
\multiput(303,547)(20.136,-5.034){0}{\usebox{\plotpoint}}
\multiput(311,545)(20.136,-5.034){0}{\usebox{\plotpoint}}
\put(322.33,542.17){\usebox{\plotpoint}}
\multiput(327,541)(20.629,-2.292){0}{\usebox{\plotpoint}}
\put(342.83,539.15){\usebox{\plotpoint}}
\multiput(344,539)(20.136,-5.034){0}{\usebox{\plotpoint}}
\multiput(352,537)(20.595,-2.574){0}{\usebox{\plotpoint}}
\put(363.25,535.64){\usebox{\plotpoint}}
\multiput(369,535)(20.595,-2.574){0}{\usebox{\plotpoint}}
\put(383.90,534.00){\usebox{\plotpoint}}
\multiput(385,534)(20.595,-2.574){0}{\usebox{\plotpoint}}
\multiput(393,533)(20.756,0.000){0}{\usebox{\plotpoint}}
\put(404.58,532.68){\usebox{\plotpoint}}
\multiput(410,532)(20.756,0.000){0}{\usebox{\plotpoint}}
\put(425.29,532.00){\usebox{\plotpoint}}
\multiput(427,532)(20.756,0.000){0}{\usebox{\plotpoint}}
\multiput(435,532)(20.756,0.000){0}{\usebox{\plotpoint}}
\put(446.05,532.00){\usebox{\plotpoint}}
\multiput(451,532)(20.756,0.000){0}{\usebox{\plotpoint}}
\put(466.80,532.00){\usebox{\plotpoint}}
\multiput(468,532)(20.756,0.000){0}{\usebox{\plotpoint}}
\multiput(476,532)(20.595,2.574){0}{\usebox{\plotpoint}}
\put(487.49,533.00){\usebox{\plotpoint}}
\multiput(493,533)(20.756,0.000){0}{\usebox{\plotpoint}}
\put(508.25,533.00){\usebox{\plotpoint}}
\multiput(509,533)(20.595,2.574){0}{\usebox{\plotpoint}}
\multiput(517,534)(20.756,0.000){0}{\usebox{\plotpoint}}
\put(528.94,534.00){\usebox{\plotpoint}}
\multiput(534,534)(20.756,0.000){0}{\usebox{\plotpoint}}
\put(549.70,534.00){\usebox{\plotpoint}}
\multiput(550,534)(20.756,0.000){0}{\usebox{\plotpoint}}
\multiput(559,534)(20.756,0.000){0}{\usebox{\plotpoint}}
\put(570.45,534.00){\usebox{\plotpoint}}
\multiput(575,534)(20.756,0.000){0}{\usebox{\plotpoint}}
\put(591.21,534.00){\usebox{\plotpoint}}
\multiput(592,534)(20.595,-2.574){0}{\usebox{\plotpoint}}
\multiput(600,533)(20.756,0.000){0}{\usebox{\plotpoint}}
\put(611.87,532.52){\usebox{\plotpoint}}
\multiput(616,532)(20.756,0.000){0}{\usebox{\plotpoint}}
\put(632.54,531.06){\usebox{\plotpoint}}
\multiput(633,531)(20.595,-2.574){0}{\usebox{\plotpoint}}
\multiput(641,530)(20.629,-2.292){0}{\usebox{\plotpoint}}
\put(653.15,528.61){\usebox{\plotpoint}}
\multiput(658,528)(20.595,-2.574){0}{\usebox{\plotpoint}}
\put(673.74,526.03){\usebox{\plotpoint}}
\multiput(674,526)(20.261,-4.503){0}{\usebox{\plotpoint}}
\multiput(683,524)(20.595,-2.574){0}{\usebox{\plotpoint}}
\put(694.12,522.22){\usebox{\plotpoint}}
\multiput(699,521)(20.136,-5.034){0}{\usebox{\plotpoint}}
\put(714.30,517.38){\usebox{\plotpoint}}
\multiput(716,517)(20.136,-5.034){0}{\usebox{\plotpoint}}
\multiput(724,515)(19.434,-7.288){0}{\usebox{\plotpoint}}
\put(734.08,511.22){\usebox{\plotpoint}}
\multiput(740,509)(19.690,-6.563){0}{\usebox{\plotpoint}}
\put(753.63,504.26){\usebox{\plotpoint}}
\multiput(757,503)(18.564,-9.282){0}{\usebox{\plotpoint}}
\put(772.35,495.33){\usebox{\plotpoint}}
\multiput(773,495)(18.144,-10.080){0}{\usebox{\plotpoint}}
\multiput(782,490)(17.601,-11.000){0}{\usebox{\plotpoint}}
\put(790.25,484.84){\usebox{\plotpoint}}
\multiput(798,480)(16.604,-12.453){0}{\usebox{\plotpoint}}
\put(807.35,473.10){\usebox{\plotpoint}}
\multiput(815,468)(15.620,-13.668){0}{\usebox{\plotpoint}}
\put(823.70,460.39){\usebox{\plotpoint}}
\put(839.26,446.66){\usebox{\plotpoint}}
\multiput(840,446)(13.789,-15.513){0}{\usebox{\plotpoint}}
\put(853.13,431.23){\usebox{\plotpoint}}
\multiput(856,428)(12.966,-16.207){0}{\usebox{\plotpoint}}
\put(866.30,415.19){\usebox{\plotpoint}}
\put(878.98,398.77){\usebox{\plotpoint}}
\multiput(881,396)(10.878,-17.677){0}{\usebox{\plotpoint}}
\put(890.02,381.21){\usebox{\plotpoint}}
\put(900.45,363.26){\usebox{\plotpoint}}
\put(910.24,344.99){\usebox{\plotpoint}}
\put(918.63,326.01){\usebox{\plotpoint}}
\put(926.30,306.72){\usebox{\plotpoint}}
\put(933.77,287.36){\usebox{\plotpoint}}
\multiput(939,274)(5.896,-19.900){2}{\usebox{\plotpoint}}
\put(952.25,227.95){\usebox{\plotpoint}}
\multiput(955,218)(4.754,-20.204){2}{\usebox{\plotpoint}}
\put(963,184){\usebox{\plotpoint}}
\sbox{\plotpoint}{\rule[-0.200pt]{0.400pt}{0.400pt}}%
\put(468,330){\makebox(0,0)[r]{$\epsilon=1$}}
\multiput(490,330)(20.756,0.000){4}{\usebox{\plotpoint}}
\put(556,330){\usebox{\plotpoint}}
\put(220,605){\usebox{\plotpoint}}
\put(220.00,605.00){\usebox{\plotpoint}}
\multiput(228,601)(20.261,-4.503){0}{\usebox{\plotpoint}}
\put(239.43,598.09){\usebox{\plotpoint}}
\multiput(245,596)(20.136,-5.034){0}{\usebox{\plotpoint}}
\put(259.51,593.19){\usebox{\plotpoint}}
\multiput(261,593)(20.261,-4.503){0}{\usebox{\plotpoint}}
\multiput(270,591)(20.595,-2.574){0}{\usebox{\plotpoint}}
\put(279.95,589.76){\usebox{\plotpoint}}
\multiput(286,589)(20.136,-5.034){0}{\usebox{\plotpoint}}
\put(300.38,586.29){\usebox{\plotpoint}}
\multiput(303,586)(20.595,-2.574){0}{\usebox{\plotpoint}}
\multiput(311,585)(20.756,0.000){0}{\usebox{\plotpoint}}
\put(321.04,584.75){\usebox{\plotpoint}}
\multiput(327,584)(20.629,-2.292){0}{\usebox{\plotpoint}}
\put(341.65,582.29){\usebox{\plotpoint}}
\multiput(344,582)(20.595,-2.574){0}{\usebox{\plotpoint}}
\multiput(352,581)(20.756,0.000){0}{\usebox{\plotpoint}}
\put(362.31,580.74){\usebox{\plotpoint}}
\multiput(369,580)(20.595,-2.574){0}{\usebox{\plotpoint}}
\put(382.91,578.26){\usebox{\plotpoint}}
\multiput(385,578)(20.756,0.000){0}{\usebox{\plotpoint}}
\multiput(393,578)(20.629,-2.292){0}{\usebox{\plotpoint}}
\put(403.59,576.80){\usebox{\plotpoint}}
\multiput(410,576)(20.595,-2.574){0}{\usebox{\plotpoint}}
\put(424.23,575.00){\usebox{\plotpoint}}
\multiput(427,575)(20.595,-2.574){0}{\usebox{\plotpoint}}
\multiput(435,574)(20.595,-2.574){0}{\usebox{\plotpoint}}
\put(444.85,572.77){\usebox{\plotpoint}}
\multiput(451,572)(20.629,-2.292){0}{\usebox{\plotpoint}}
\put(465.50,571.00){\usebox{\plotpoint}}
\multiput(468,571)(20.595,-2.574){0}{\usebox{\plotpoint}}
\multiput(476,570)(20.595,-2.574){0}{\usebox{\plotpoint}}
\put(486.12,568.76){\usebox{\plotpoint}}
\multiput(493,568)(20.756,0.000){0}{\usebox{\plotpoint}}
\put(506.78,567.28){\usebox{\plotpoint}}
\multiput(509,567)(20.595,-2.574){0}{\usebox{\plotpoint}}
\multiput(517,566)(20.756,0.000){0}{\usebox{\plotpoint}}
\put(527.45,565.82){\usebox{\plotpoint}}
\multiput(534,565)(20.595,-2.574){0}{\usebox{\plotpoint}}
\put(548.04,563.24){\usebox{\plotpoint}}
\multiput(550,563)(20.756,0.000){0}{\usebox{\plotpoint}}
\multiput(559,563)(20.595,-2.574){0}{\usebox{\plotpoint}}
\put(568.71,561.79){\usebox{\plotpoint}}
\multiput(575,561)(20.756,0.000){0}{\usebox{\plotpoint}}
\put(589.38,560.29){\usebox{\plotpoint}}
\multiput(592,560)(20.595,-2.574){0}{\usebox{\plotpoint}}
\multiput(600,559)(20.756,0.000){0}{\usebox{\plotpoint}}
\put(610.04,558.75){\usebox{\plotpoint}}
\multiput(616,558)(20.629,-2.292){0}{\usebox{\plotpoint}}
\put(630.69,557.00){\usebox{\plotpoint}}
\multiput(633,557)(20.595,-2.574){0}{\usebox{\plotpoint}}
\multiput(641,556)(20.629,-2.292){0}{\usebox{\plotpoint}}
\put(651.32,554.84){\usebox{\plotpoint}}
\multiput(658,554)(20.756,0.000){0}{\usebox{\plotpoint}}
\put(671.98,553.25){\usebox{\plotpoint}}
\multiput(674,553)(20.629,-2.292){0}{\usebox{\plotpoint}}
\multiput(683,552)(20.595,-2.574){0}{\usebox{\plotpoint}}
\put(692.59,550.80){\usebox{\plotpoint}}
\multiput(699,550)(20.595,-2.574){0}{\usebox{\plotpoint}}
\put(713.19,548.31){\usebox{\plotpoint}}
\multiput(716,548)(20.595,-2.574){0}{\usebox{\plotpoint}}
\multiput(724,547)(20.595,-2.574){0}{\usebox{\plotpoint}}
\put(733.79,545.78){\usebox{\plotpoint}}
\multiput(740,545)(20.629,-2.292){0}{\usebox{\plotpoint}}
\put(754.40,543.33){\usebox{\plotpoint}}
\multiput(757,543)(20.595,-2.574){0}{\usebox{\plotpoint}}
\multiput(765,542)(20.595,-2.574){0}{\usebox{\plotpoint}}
\put(775.00,540.78){\usebox{\plotpoint}}
\multiput(782,540)(20.136,-5.034){0}{\usebox{\plotpoint}}
\put(795.42,537.32){\usebox{\plotpoint}}
\multiput(798,537)(20.595,-2.574){0}{\usebox{\plotpoint}}
\multiput(806,536)(20.261,-4.503){0}{\usebox{\plotpoint}}
\put(815.87,533.89){\usebox{\plotpoint}}
\multiput(823,533)(20.595,-2.574){0}{\usebox{\plotpoint}}
\put(836.38,530.81){\usebox{\plotpoint}}
\multiput(840,530)(20.595,-2.574){0}{\usebox{\plotpoint}}
\multiput(848,529)(20.136,-5.034){0}{\usebox{\plotpoint}}
\put(856.71,526.82){\usebox{\plotpoint}}
\multiput(864,525)(20.629,-2.292){0}{\usebox{\plotpoint}}
\put(877.06,522.98){\usebox{\plotpoint}}
\multiput(881,522)(20.136,-5.034){0}{\usebox{\plotpoint}}
\multiput(889,520)(20.136,-5.034){0}{\usebox{\plotpoint}}
\put(897.20,517.96){\usebox{\plotpoint}}
\multiput(906,516)(19.434,-7.288){0}{\usebox{\plotpoint}}
\put(917.10,512.22){\usebox{\plotpoint}}
\multiput(922,511)(19.434,-7.288){0}{\usebox{\plotpoint}}
\put(936.79,505.74){\usebox{\plotpoint}}
\multiput(939,505)(18.564,-9.282){0}{\usebox{\plotpoint}}
\multiput(947,501)(19.434,-7.288){0}{\usebox{\plotpoint}}
\put(955.84,497.58){\usebox{\plotpoint}}
\put(963,494){\usebox{\plotpoint}}
\end{picture}

\caption{The function $f(\epsilon,\nu)$ vs. $\nu$ at several characteristic
values of energy: low energy, no secondary maximum ($\epsilon=1$),
the secondary maximum just developed ($\epsilon=10$), the secondary maximum
becomes global and is just above the unitarity limit ($\epsilon=15.5$)[80].}
\end{center} \end{figure}

\subsection{Generating field technique.}
\subsubsection{Tree level.}
A more convenient and more conceptually transparent technique for dealing
with tree-level threshold multiboson amplitudes was suggested by
Brown\cite{brown} and was later extended to calculation of
one-loop\cite{mv9,smith} and higher quantum effects\cite{mv10,gv} in
these amplitudes. The technique is based on the standard
reduction formula representation of the amplitude through the response of
the system to an external source $\rho(x)$, which enters the term $\rho
\phi$ added to the Lagrangian.
\begin{eqnarray}
&&\langle n|\phi(x)|0\rangle  =
\left [ \prod_{a=1}^n \lim_{p_a^2 \to m^2}
\int d^4 x_a~ e^{i p_a x_a} (m^2- \right . \nonumber \\
&& \left . p_a^2)~ {\delta \over {\delta \rho(x_a)}}
\right ] \left . \langle 0_{out}|\phi(x)|0_{in}\rangle ^\rho
 \right|_{\rho=0}~~,
\label{reduc}
\end{eqnarray}
the tree-level amplitude being generated by
the response in the classical approximation, i.e. by the classical solution
$\phi_0(x)$ of the field equations in the presence of the source.

For all the spatial momenta of the final particles equal to zero it is
sufficient to consider the response to a spatially uniform time-dependent
source $\rho(t) = \rho_0(\omega)\, e^{i\omega t}$ and take the on-mass-shell
limit in eq.(\ref{reduc}) by tending $\omega$ to $m$. The spatial integrals
in eq.(\ref{reduc}) then give the usual factors with the normalization
spatial volume, which as usual is set to one, while the time dependence on
one common frequency $\omega$ implies that the propagator factors and the
functional derivatives enter in the combination
\beq
(m^2-p_a^2){\delta \over {\delta \rho(x_a)}} \to (m^2-\omega^2) {\delta
\over {\delta \rho(t)}}={\delta \over {\delta z(t)}} \label{zt}~~, \eeq
where
\beq
z(t)={{\rho_0(\omega)\, e^{i \omega t}} \over {m^2-i \epsilon -\omega^2}}
\label{zl}~~
\eeq
coincides with the response of the field to the external source in the limit
of absence of the interaction, i.e. of $\lambda =0$. For a finite amplitude
$\rho_0$ of the source the response $z(t)$ is singular in the limit $\omega
\to m$. The crucial observation of Brown\cite{brown} is that, since
according to eq.(\ref{zt}) we need the dependence of the response of the
interacting field $\phi$ only in terms of $z(t)$, one can take the limit
$\rho_0(\omega) \to 0$ simultaneously with $\omega \to m$ in such a way that
$z(t)$ is finite:
$z(t) \to z_0 e^{imt}$.

Furthermore, to find the classical solution $\phi_0(x)$ in this limit one
does not have to go through this limiting procedure, but rather consider
directly the on-shell limit with vanishing source. The field equation with
zero source in $\lambda \phi^4$ theory without SSB is
\beq
\partial^2 \phi + m^2 \phi + \lambda \phi^3 =0~~.
\label{el}
\eeq
For the purpose of calculating the matrix element in eq.(\ref{reduc}) at the
threshold one looks for a solution of this equation which depends only on
time and contains only the positive frequency part with all harmonics being
multiples of $e^{imt}$, which condition is equivalent to requiring that
$\phi(t) \to 0$ as Im$\,t \to +\infty$.
The solution satisfying these conditions reads as\cite{brown}
\beq
\phi_0(t)={{z(t)} \over {1-(\lambda/8 m^2) z(t)^2}}~~
\label{phi0}
\eeq

According to equations (\ref{zt}) and (\ref{reduc}) the $n$-th derivative of
this solution with respect to $z$ gives the matrix element $\langle
n|\phi(0)|0\rangle $ at the threshold in the tree approximation:
\beq
\langle 2k+1|\phi(0)|0\rangle _0=
\left . \left ( {\partial \over {\partial z}}
\right )^{2k+1} \phi_0 \right |_{z=0}~~,
\label{a0}
\eeq
which reproduces the result in eq.(\ref{sol1}).

For the case of theory with SSB the solution reads as
\beq
\phi_0(t)=-v{{1+z/2v} \over {1-z/2v}}~,
\label{phi01}
\eeq
which reproduces the tree amplitudes in eq.(\ref{sol2}). In this case
$z(t)=e^{iMt}$, where $M=\sqrt{2}\,|m|$ is the mass of physical scalar
boson.

\unitlength 1mm
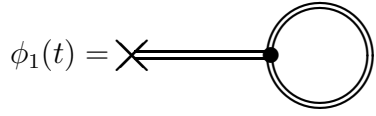
\begin{figure} \begin{center}
\begin{picture}(62.00,17.00)
\put(55.00,10.00){\circle{14.00}}
\put(55.00,10.00){\circle{12.00}}
\put(30.36,9.50){\line(1,0){18.00}}
\put(30.36,10.50){\line(1,0){18.00}}
\put(28.00,12.00){\line(1,-1){4.00}}
\put(28.00,8.00){\line(1,1){4.00}}
\put(48.50,10.00){\circle*{2.00}}
\put(27.00,10.00){\makebox(0,0)[rc]{$\phi_1(t)=$}}
\end{picture}
\caption{The tadpole graph in the background  field $\phi_0$ for calculating
one-loop correction to $\phi_0$. The double lines and the heavy dot
represent respectively the Green's function and the vertex in the background
field.}
\end{center} \end{figure}

\subsubsection{One-loop level.}
To advance the calculation to the one-loop level one has to calculate the
first quantum correction $\phi_1(t)$ to the classical background field
$\phi_0$. This amounts\cite{mv9} to evaluating the tadpole graph of figure
6, where both the Green's function and the vertex are calculated in the
external background field $\phi_0$. The green's function $G(x;\,x^\prime)$
satisfies the equation
\beq
\left ( \partial^2 + m^2 + 3\lambda\,\phi_0(t)^2 \right )G(x;\,x^\prime)
=-i\,\delta(x-x^\prime)~,
\label{eqgreen}
\eeq
in which the differential operator in the Minkowski time contains explicitly
complex field $\phi_0$ (cf. eq.(\ref{phi0}) or eq.(\ref{phi01})). A
straightforward rotation to the Euclidean time, $i\,t \to \tau$, is
problematic, since the background field then develops a pole at a real
$\tau$. The acceptable solution is achieved by simultaneously rotating and
shifting the time axis in eq.(\ref{eqgreen}) in such a way that
$-\lambda \,z(t)^2/8m^2 \to \exp(2m\tau)$ for the theory without SSB, and
$-z(t)/2v \to \exp(M\tau)$ for the theory with SSB. In terms of thus defined
$\tau$ the background field has the form $\phi_0(\tau)=i\,\sqrt{2/\lambda}\,
m / \cosh(m \tau)$ (no SSB) and $\phi_0(\tau)=v\, \tanh(M \tau/2)$ (with
SSB). In both cases the term $\phi_0(t)^2$ in equation (\ref{eqgreen}) is
real and non-singular. After applying the standard decomposition of the
Green's function over the conserved in the background $\phi_0(t)$ spatial
momentum ${\bf k}$:
\beq
G(\tau,\,{\bf x};\, \tau^\prime,\,{\bf x}^\prime)= \int G_\omega(\tau,\,
\tau^\prime) \, e^{i {\bf k} ({\bf x}-{\bf x}^\prime)}\, {{d^{d-1}k} \over
{(2\pi)^{d-1}}}~,
\label{fourier}
\eeq
one arrives for the case of no SSB
at the well-known in Quantum Mechanics equation
\beq
\left ( - {d^2 \over d\tau^2}+\omega^2 - {6 \over {(\cosh \tau)^2}} \right
)=\delta(\tau-\tau^\prime)
\label{qmeq}
\eeq
with $\omega=\sqrt{{\bf k}^2+1}$ and the mass $m$ set to one.
(For the theory with SSB one gets the same equation with a rescaled
$\omega$\cite{smith}.) Thus the problem of finding the first quantum
correction $\phi_1(t)$ to the background field is completely solvable on the
$\tau$ axis, and the solution can then be extended to the whole complex
plane of $t$ by analytical continuation. For the theory with no SSB the
result\cite{mv9} for the amplitudes $a(n)$ at the one-loop level reads as
\beq
a(n)_{0+1}=a(n)_0 \left (1-(n-1)(n-3)\, {3\lambda \over 32}\, F \right )
\label{an01}~,
\eeq
where
\beq
F={\sqrt{3} \over 2\pi^2} \left ( \ln {{2+\sqrt{3}} \over {2-\sqrt{3}}} -
i\pi \right )~.
\label{F}
\eeq
The analog of this result for the SSB case is\cite{smith}
\beq
a(n)_{0+1}=a(n)_0 \left (1+n(n-1)\,{\sqrt{3}\,\lambda \over 8\pi} \right )~.
\label{an011}
\eeq

\subsection{Nullification of threshold amplitudes.}
The equations (\ref{an01}) and (\ref{an011}) display a remarkable feature:
in spite of the presence of an intermediate state with two bosons in
one-loop graphs, their contribution to the amplitudes in the case of SSB is
real, while the factor $F$ in eq.(\ref{F}) is an easily recognizable
threshold factor for the $2 \to 4$ process with no indication of presence of
other thresholds. Using the unitarity relation for the imaginary part of the
loop graphs, one immediately concludes that this can only be if the tree
amplitudes of the on-shell processes $2 \to n$ are all zero at the threshold
for $n > 4$ in the theory without SSB\cite{mv9,mv11} and for $n > 2$ in the
theory with SSB\cite{smith}.

This can be traced to
the special properties of the reflectionless potential $-6/(\cosh\tau)^2$ in
equation (\ref{qmeq}) and generalized\cite{mv12,akp4} to other theories,
where the problem of the $2 \to n$ scattering is reduced to finding the
Green's function in the reflectionless potential $-N (N+1)/(\cosh\tau)^2$
with integer $N$.
The known additional cases are the following:
\begin{itemize}
\item{Linear $\sigma$ model ($N=1$): the tree-level amplitudes of the
scattering $\pi \, \pi \to n\,\sigma$ are all zero at the corresponding
thresholds for $n > 1\,$.}
\item{Fermions with Higgs-generated mass: if $2m_f/m_H = N$ (integer), then
all tree-level amplitudes of $f\,{\overline f} \to n\,H$ are zero at
threshold for all $n \ge N$.}
\item{Vector bosons with Higgs-generated mass: if $4 m_V^2/m_H^2 =N(N+1)$,
then all tree-level amplitudes with transversal vector bosons of $V_T\,V_T
\to n\, H$ are zero at threshold for all $n > N$. For longitudinal vectors
the same amplitudes of $V_L\,V_L \to n \, H$ are zero for
$n=N\,$\cite{smith2} and for $n > N+1\,$\cite{akp4}.}
\end{itemize}
All these cases, except the one with longitudinal vectors,
stem\cite{mv12} from the generic interaction of two fields $\phi$ and $\chi$
of the form ${\xi \over 2} \, \phi^2\,\chi^2$ and the self-interaction of
the field $\phi$ as described by the potential (\ref{phi4}). Then if
the ratio of the coupling constants satisfies the relation $2
\xi/\lambda =N(N+1)$ with $N$ integer, the tree-level threshold amplitudes
of the processes $2 \chi \to n\, \phi$ are all zero for $n > N$ in a theory
with SSB and for $n > 2 N$ in a theory with unbroken symmetry. This behavior
somewhat resembles the nullification of inelastic amplitudes in the
Sine-Gordon theory, where it is a consequence of a symmetry and is a deep
property of the theory. In the theories considered here this is a much
weaker property, which holds only at threshold and, generally, only at the
tree level\cite{smith3}. However the nullification in this case can be a
consequence of a hidden symmetry, which holds at the classical level and/or
has a more restricted scope. Thus far such symmetry has been
revealed\cite{lrt} only for the case of $N=1$, where it can be traced to the
symmetry of a system of two anharmonic oscillators, described by the
potential
\beq
V(x,y)={\omega_1^2 \over 2} x^2 +{\omega_2^2 \over 2} y^2 + {\lambda \over
4} (x^2+y^2)^2 ~.
\label{vxy}
\eeq
If the frequencies $\omega_1$ and $\omega_2$ were equal, the model would
have an O(2) symmetry, corresponding to conservation of the angular momentum
$Q={\dot x} y - x {\dot y}$. However even for $\omega_1 \ne \omega_2$ the
symmetry persists\cite{lrt} corresponding to conservation of the invariant
${\lambda \over 4} Q^2 + (\omega_1^2-\omega_2^2) \bigl ({1 \over 2}
{\dot y}^2+{\omega_2^2 \over 2}y^2+{\lambda \over 4} y^4 + {\lambda \over 4}
x^2 y^2 \bigr )$.

It should be also noted that if the ratio $2\xi/\lambda$ does not satisfy
the above mentioned condition, the threshold amplitudes of the processes $2
\chi \to n\, \phi$ display\cite{bz} a `normal' factorial growth with $n$.

\subsection{Non-perturbative analysis.}

The $n^2 \lambda$ behavior of the loop corrections in the equations
(\ref{an01}) and (\ref{an011}) convinces us that the perturbation theory is
of little help in finding the amplitudes at large $n$ and a true
non-perturbative analysis is required. It turns out that to a certain extent
such analysis can be performed for the $\lambda \phi^4$ theory with SSB.
In terms of the variable $\tau$ the problem of calculating the threshold
amplitudes $a(n)=\langle n |\phi(0)|0 \rangle$ reduces\cite{mv10} to a well
defined Euclidean-space problem of calculating the quantum average
$\Phi(\tau)$ of the field
\beq
\Phi(\tau)={{\int \left ( \int \phi(\tau,\,{\bf x}) \, d{\bf x} /
\int d{\bf x} \right ) \, e^{-S[\phi]}\, {\cal D}\phi} \over
{\int e^{-S[\phi]} \, {\cal D}\phi}}
\label{qphi}
\eeq
with the kink boundary conditions, i.e. $\phi \to \pm v$ as $\tau \to \pm
\infty$. The average field then expands at $\tau \to -\infty$ in the series
\beq
\Phi(\tau)=\sum_{n=0}^\infty c_n \, e^{nM\tau}
\label{expand}
\eeq
and the threshold amplitudes are given by $a(n)=n! \, c_n/c_1^n$, where the
coefficient $c_1$ describes the one-particle state normalization:
$c_1=\langle 1 |\phi(0)|0 \rangle$.

Due to the fact that the classical kink solution provides the absolute
minimum for the action under specified boundary conditions, the path
integrals in eq.(\ref{qphi}) are well defined (no negative modes) and thus
the average field $\Phi(\tau)$ is real in any finite order of the perturbation
theory. Thus the coefficients $c_n$ of the expansion (\ref{expand}) are real
too. Therefore the amplitudes $a(n)$ are real to any finite order in
$\lambda$. As we have seen at the one loop level this implies the
nullification of the tree amplitudes of $2 \to n$ for $n > 2$. In higher
loops this implies a relation between the amplitudes of the processes
$k \to n$ with different $k$. The only exception is the particular case of
$n=3$, for which the imaginary part of the $a(3)$ can be contributed only by
the two-boson intermediate state. Since the imaginary part is vanishing,
one concludes\cite{mv10} that the $2 \to 3$ amplitude is vanishing at the
threshold in all orders in $\lambda$.

The function $\Phi(\tau)$ given by the expansion (\ref{expand}) is
manifestly periodic: $\Phi(\tau + 2i\pi/m)=\Phi(\tau)$. Using this property
and the  boundary conditions at $\tau \to \pm \infty$ one finds\cite{mv10}
that the exact function $\Phi(\tau)$ necessarily has a singularity at a
finite $\tau$. Thus the expansion (\ref{expand}) has a finite radius of
convergence, and thus the exact threshold amplitudes $a(n)$ grow at least as
fast as $n!$. In other words, the quantum effects  do
not eliminate the factorial growth of $a(n)$.

As is indicated by the $n^2 \lambda$ parameter of the perturbation theory
for the coefficients $c_n$, the saddle point (SP) for the action $S[\phi]$,
given by the ${\bf x}$-independent `domain wall' with the kink profile is
not the correct SP for calculating the coefficients $c_n$ at large $n$. It
has been argued\cite{gv} that the correct SP configuration is given by a
spatially inhomogeneous field configurations in which the `domain wall' is
deflected towards negative $\tau$ by a maximal amount $h_0$. Then at a large
negative $\tau$ the coefficients $c_n$ are given by
\beq
c_n \sim e^{n M h_0} \, e^{-\mu (A-A_0)}~,
\label{seff}
\eeq
where $A$ is the area of the domain wall with deflection, $A_0$ is the same
for the undistorted flat wall, and $\mu = |m|^3/3\lambda$ is the surface
tension of the wall.

Finding the extremum of the `effective action' in the
exponent in eq.(\ref{seff}) exactly corresponds to finding the equilibrium
configuration of a $(d-1)$ dimensional membrane with a force equal to $nM$
applied at the point of maximum deflection. In general this problem has no
real solution (the film gets punched). However a solution exists, where
a part of the trajectory of the domain wall resides in the Euclidean space,
and a part is in the Minkowski space\cite{gv}. The Minkowski-space part of
the trajectory corresponds to evolution of a bubble made of a domain wall
and having energy $E=n\,M$. The amplitudes $a(n)$ are then found as a sum of
resonant contributions of the quantized levels of the bubble:
\beq
a(n) \sim n! {{e^{i\,I(E)/2}} \over {1-e^{i\,I(E)}}} \, \exp \left [ f(d)\,E
\, (E/\mu)^{1 \over {d-2}} \right ]~,
\label{annp}
\eeq
where $f(d)$ is a {\em positive} coefficient depending on the space-time
dimension, and $I(E)=\oint p\, dr$ is the action of the bubble over one
period of oscillation. Clearly the amplitudes $a(n)$ in eq.(\ref{annp}) have
poles at the values of $E$ satisfying the Bohr-Sommerfeld condition
$I(E)=2\pi N$.

The result in eq.(\ref{annp}) can be interpreted as that the growth of
$a(n)$ is due to a strong coupling of the bubble states to the multi-boson
states with all particles being exactly at rest. However, it is known from
the numerical studies of mid-70s[93 - 95] that the lifetime of the
bubbles is of order one in units of their period. Thus one should
conclude\cite{gv} that there arises a non-perturbative form factor, which
cuts off the integral over the phase space of the final bosons and makes the
total probability of a moderate value, inspite of the extremely large value
of the coupling to exactly static bosons. The total probability of the
process $1 \to n$ in this picture is given by the probability of creating a
bubble with energy $E$ by a virtual field $\phi$. This can be
estimated\cite{gv} by the Landau WKB method and is found to be exponentially
small:
\beq
\sigma (1 \to B(E)) \sim  \exp \left [-2 f(d)\,E
\, (E/\mu)^{1 \over {d-2}} \right ]~,
\eeq
where $f(d)$ is the same as in eq.(\ref{annp}). Thus one concludes that in
the theory with SSB the total cross section of non-perturbative
multiparticle production is extremely likely to be exponentially small at
high energy. However because of the usage of special properties of the
theory with SSB it is not clear, whether this conclusion can be generalized
to other theories, in particular, to the multi-boson production in the
Standard Model.

\section{Conclusions. Problems}
The problem of multi-particle processes in theories with weak interaction is
one of most challenging in the quantum field theory. In solving this problem
we are most likely to find new methods of non-perturbative analysis of the
field dynamics. As it stands now, there are mostly problems facing us, some
of which are:
\begin{itemize}
\item{It is not clear, to what extent the exponential suppression of the
(B+L) violation in particle collisions is lifted at high energy: by a factor
1/2 in the exponent, by a different factor, or completely. All these types
of behavior are observed in simplified models.}
\item{The $n!$ behavior of the amplitudes for production of $n$ bosons
survives the quantum effects, at least in some models. However this does not
necessarily imply a catastrophic growth of the cross section.}
\item{Peculiar zeros are observed in threshold amplitudes of multi-boson
production. However it is not clear, whether they signal some deep
properties or this is a mere coincidence.}
\item{The classical field configurations give rise to multiparticle
amplitudes. However their r\^ole in high-energy collisions is yet to be
understood.}
\end{itemize}

\vspace{6mm}

\noindent
This work and the author's participation in the conference are
supported, in part, by the DOE grant DE-AC02-83ER40105.

\vspace{20mm}

\noindent{\Large \bf Discussion}

\vspace{10mm}

\noindent{\it A.\ Kataev, CERN and INR, Moscow:}\\
Do the theoretical results discussed in your talk
have any interesting phenomenological implications?
Is it possible to study these non-perturbative effects
experimentally?\\

\noindent{\it M.\ Voloshin:}\\
Possible effects in hard processes in QCD
with production of many minijets might be observable at LHC
energies. An experimental study of possible non-perturbative processes in
the electroweak interactions may require an energy of about 1000 TeV.\\

\noindent{\it V.\ Kuvshinov, Minsk:}\\
It seems we have here new mechanisms for multiparticle
production.  For example, it can give contributions to
the intermittency phenomenon.  Is multiplicity important
for $B+L$ violation?  Or is only $n$! important? \\

\noindent{\it M.\ Voloshin:}\\
The multiplicity is important through the $n!$, or, possibly,
a stronger factor.

\end{document}